\documentclass{egpubl}
\usepackage{eurovis2020}

\usepackage[T1]{fontenc}
\usepackage{dfadobe}
\usepackage[utf8]{inputenc}
\usepackage[export]{adjustbox}

\usepackage{cite}  
\BibtexOrBiblatex
\electronicVersion
\PrintedOrElectronic

\usepackage{egweblnk}

\title{ Boxer: Interactive Comparison of Classifier Results }

\author[M. Gleicher \& A. Barve \& X. Yu \& F. Heimerl]
{\parbox{\textwidth}{\centering Michael Gleicher \orcid{0000-0003-3295-4071},
        Aditya Barve, 
        Xinyi Yu, and
        Florian Heimerl \orcid{0000-0002-3943-2260}
        }
        \\
{\parbox{\textwidth}{\centering University of Wisconsin -- Madison}}
		\\
}

\usepackage{xspace}
\usepackage{todonotes}
\usepackage{xcolor}

\newcommand{\sysname}{Boxer\xspace}

\begin{document}

\teaser{
	\vspace{-.5in}
	\centerline{\includegraphics{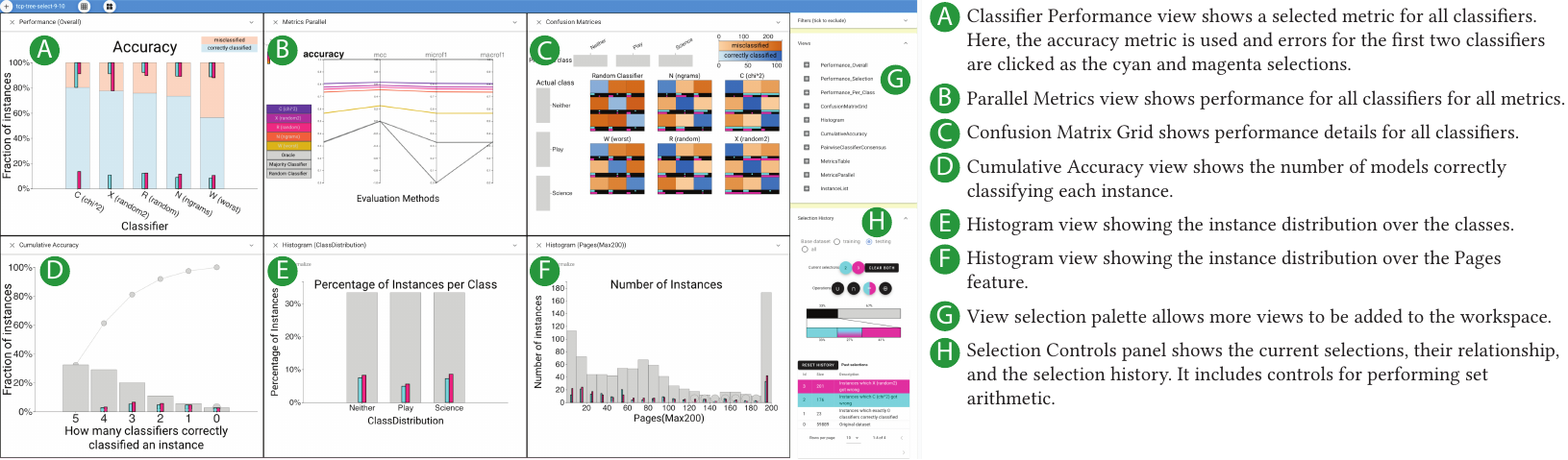}}
	\caption{\label{fig:tcp-feat}\label{fig:teaser}
		Boxer examining the results of 5 classifiers in the Feature Selection use case of Section \ref{sec:tcp}.
	}
}

\maketitle
\begin{abstract}
	Machine learning practitioners often compare the results of different classifiers to help select, diagnose and tune models. We present \emph{Boxer,} a system to enable such comparison. Our system facilitates interactive exploration of the experimental results obtained by applying multiple classifiers to a common set of model inputs. The approach focuses on allowing the user to identify interesting subsets of training and testing instances and comparing performance of the classifiers on these subsets. The system couples standard visual designs with set algebra interactions and comparative elements.  This allows the user to compose and coordinate views to specify subsets and assess classifier performance on them. The flexibility of these compositions allow the user to address a wide range of scenarios in developing and assessing classifiers. We demonstrate Boxer in use cases including model selection, tuning, fairness assessment, and data quality diagnosis.


\end{abstract}

\newcommand\differentboxes{
\begin{figure}[t]
	\centerline{\includegraphics[width=\columnwidth]{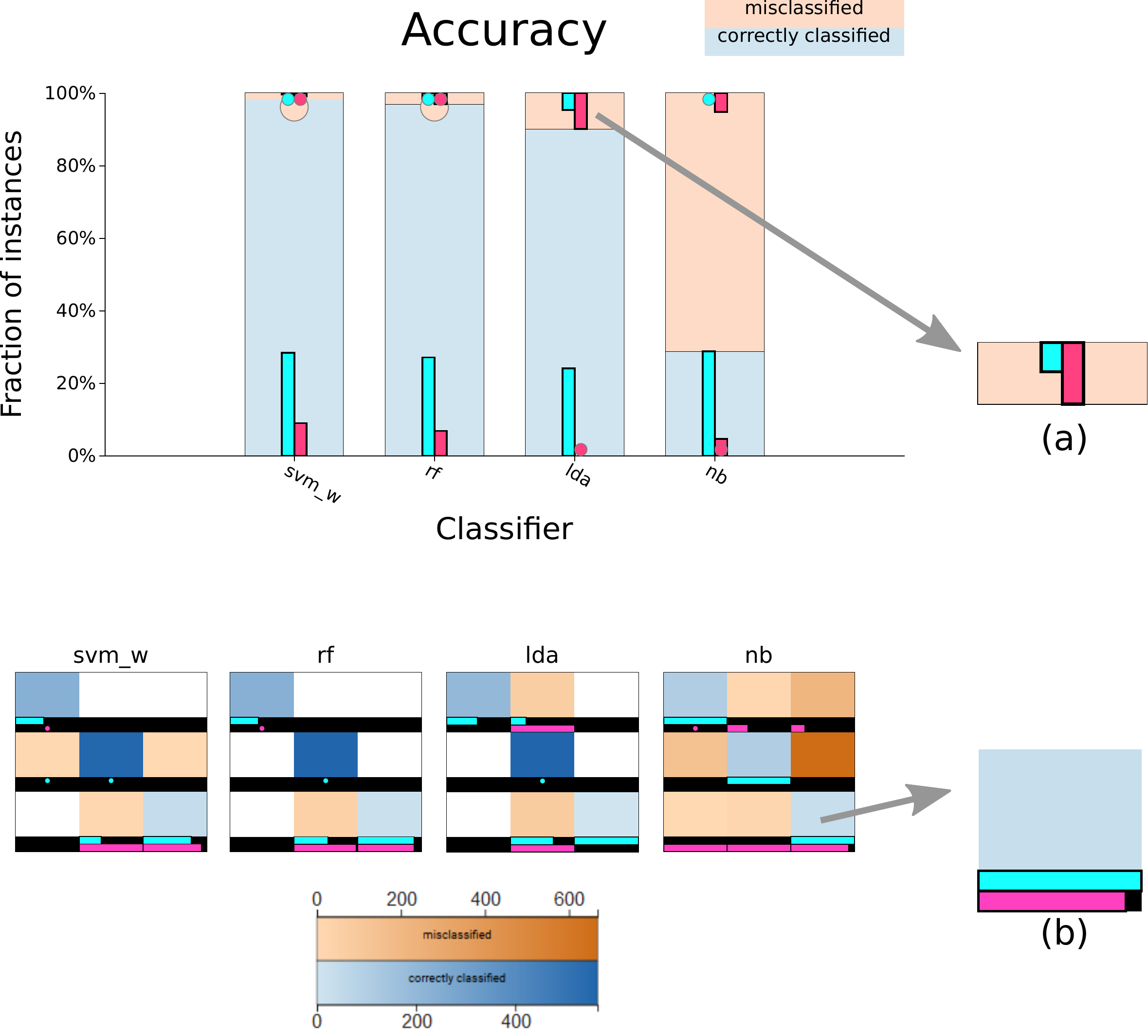}}
	\caption{\label{fig:different_boxes}
		Visualizations composed of \emph{boxes} and \emph{selections}:
		Each bar of a bar chart (a) and square of the confusion matrix (b) is a box corresponding to a subset of the data.
		All boxes show their overlaps with the current selections with color coded stripes. Dots in bars indicate small non-zero values.
	}
\end{figure}
}

\newcommand\selectioncontrolfigure{
	\begin{figure}[t]
		\centerline{\includegraphics[width=\columnwidth]{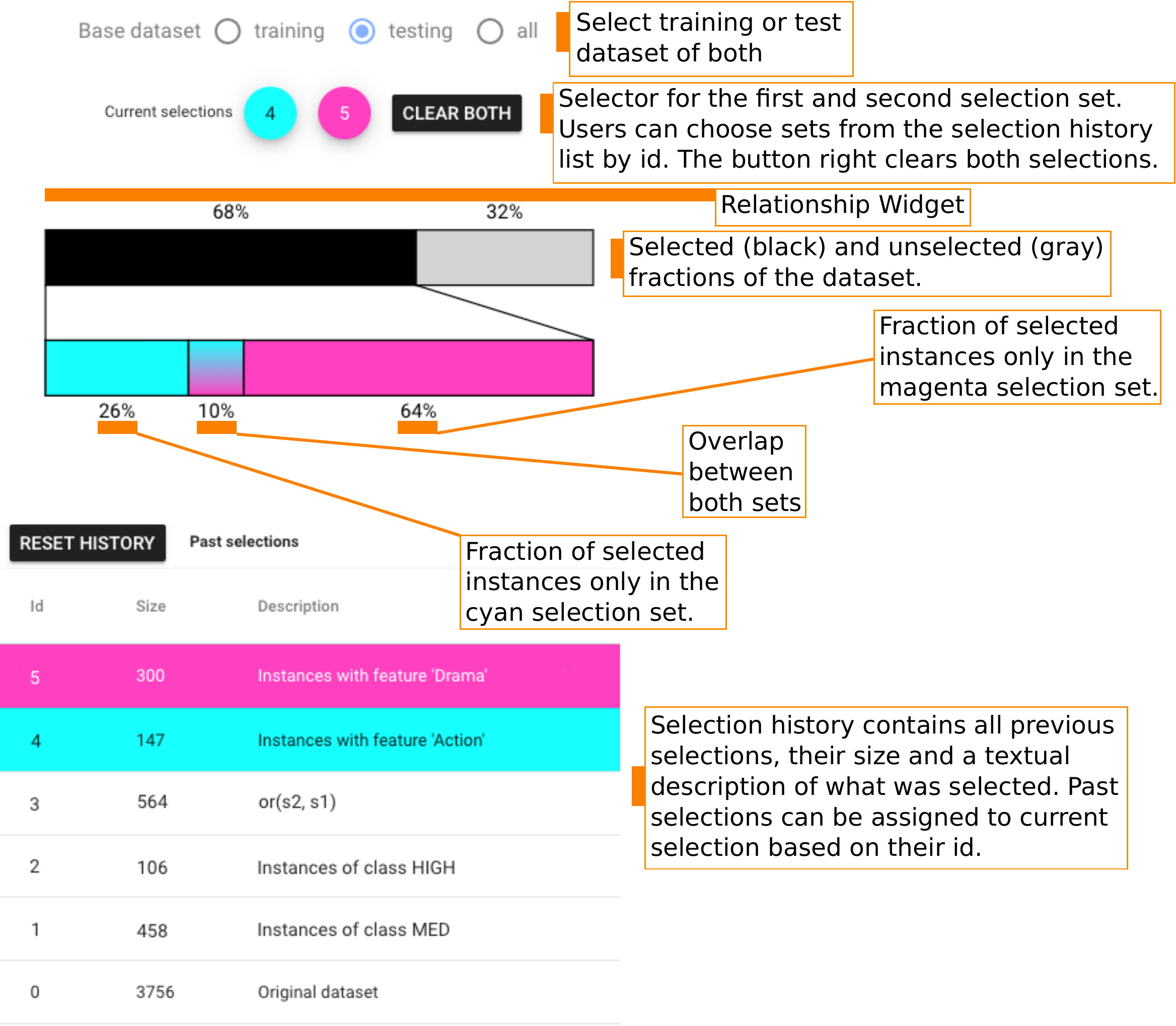}}
		\caption{\label{fig:selection_control}
		The \emph{Selection Control} view provides information about the active selections and controls to adjust them.
		The view provides a textual description of the active selections, as well as a visual indicator that relates these subsets to the overall data set and to each other.
		The lower row of the indicator shows the relation between the two active selections, including their intersection and differences.
		Clicking on the indicator selects that subset, allowing for easy set operations (e.g., clicking on the middle area to select the intersection).
		The Selection Control view provides a history list of previous selections, which can be clicked to recall one.
		}
	\end{figure}
}

\newcommand\fuzzfigure{
\begin{figure}[t]
	\centerline{
		\includegraphics[width=\columnwidth/2]{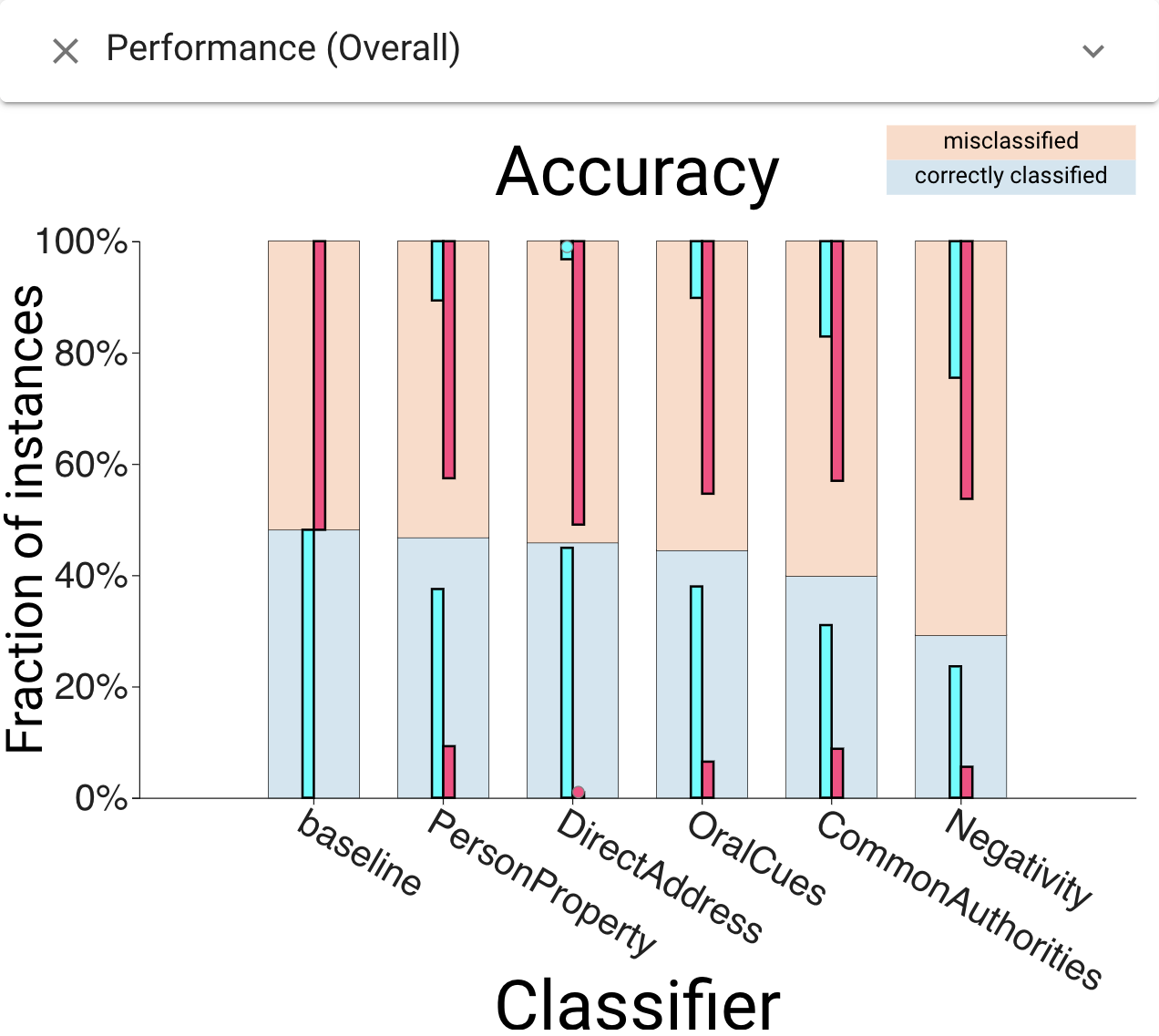}
		\includegraphics[width=\columnwidth/2]{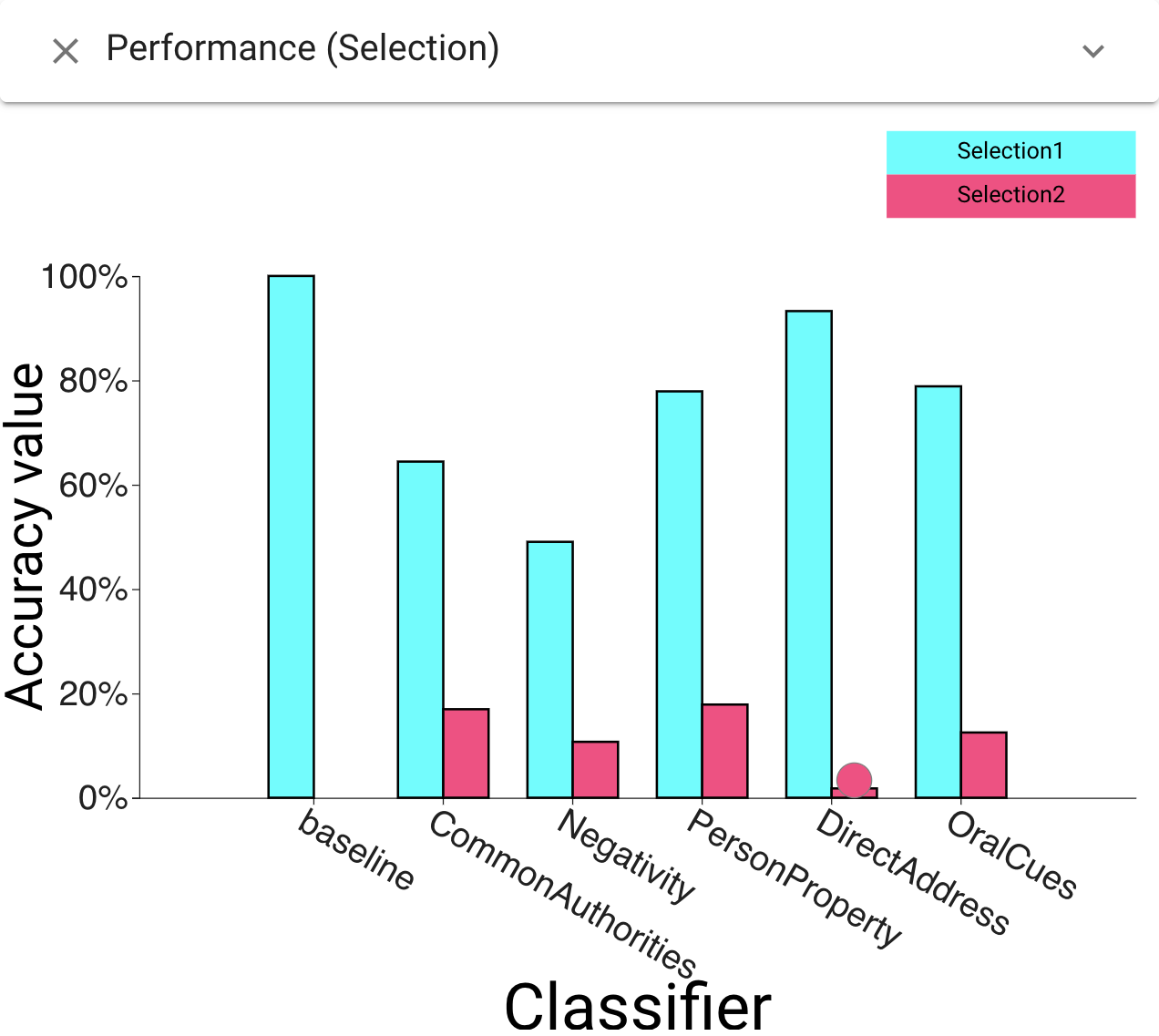}
	}
	\caption{\label{fig:fuzz}
			Sensitivity testing use case (Section~\ref{sec:fuzz}).
			The \emph{Classifier Performance} view (left) shows that (PP and DA) perform similar to the baseline. Selecting baseline correct (cyan) and incorrect (magenta) shows considerable overlap with DA, but not PP. This suggests the model is sensitive to PP, despite similar aggregate performance to the baseline. The \emph{Selection Performance} view (right) shows this as well: PP achieves 20\% accuracy on the baseline's errors.
	}
\end{figure}
}

\newcommand\imdbfigure{
	\begin{figure*}
		\centerline{
			\includegraphics[width=\textwidth/5, valign=t]{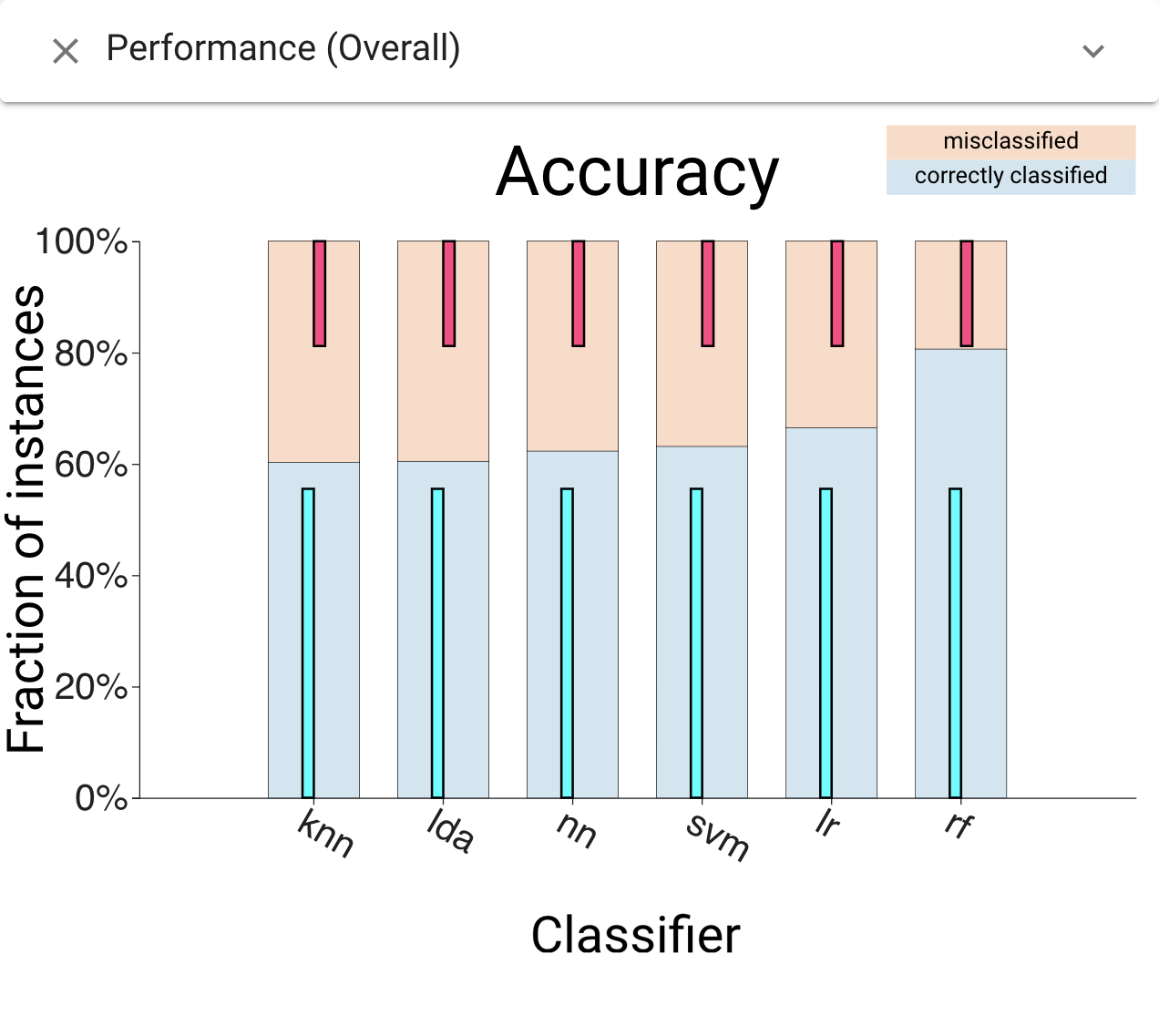}
			\includegraphics[width=\textwidth/5, valign=t]{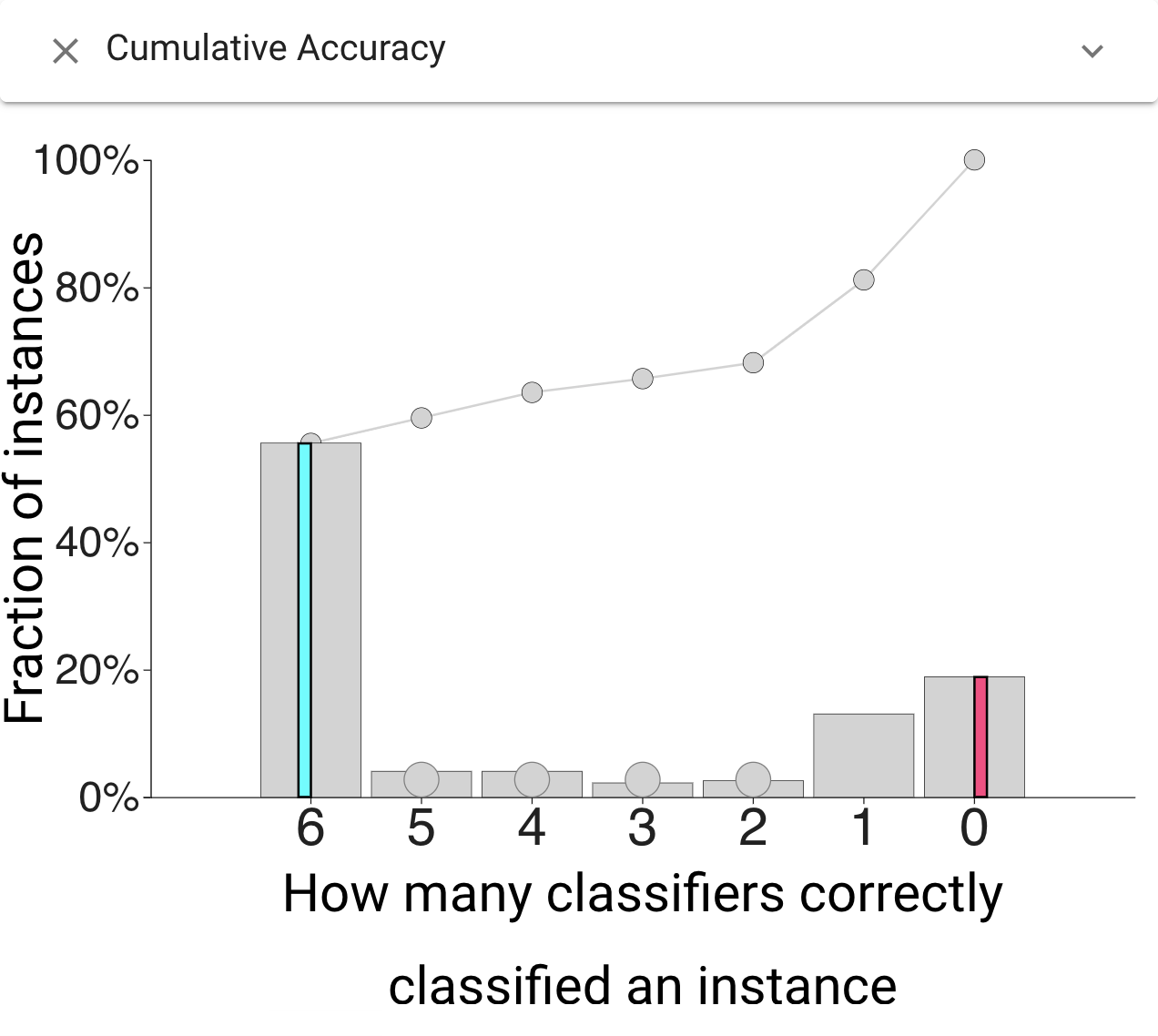}
			\includegraphics[width=\textwidth/5, valign=t]{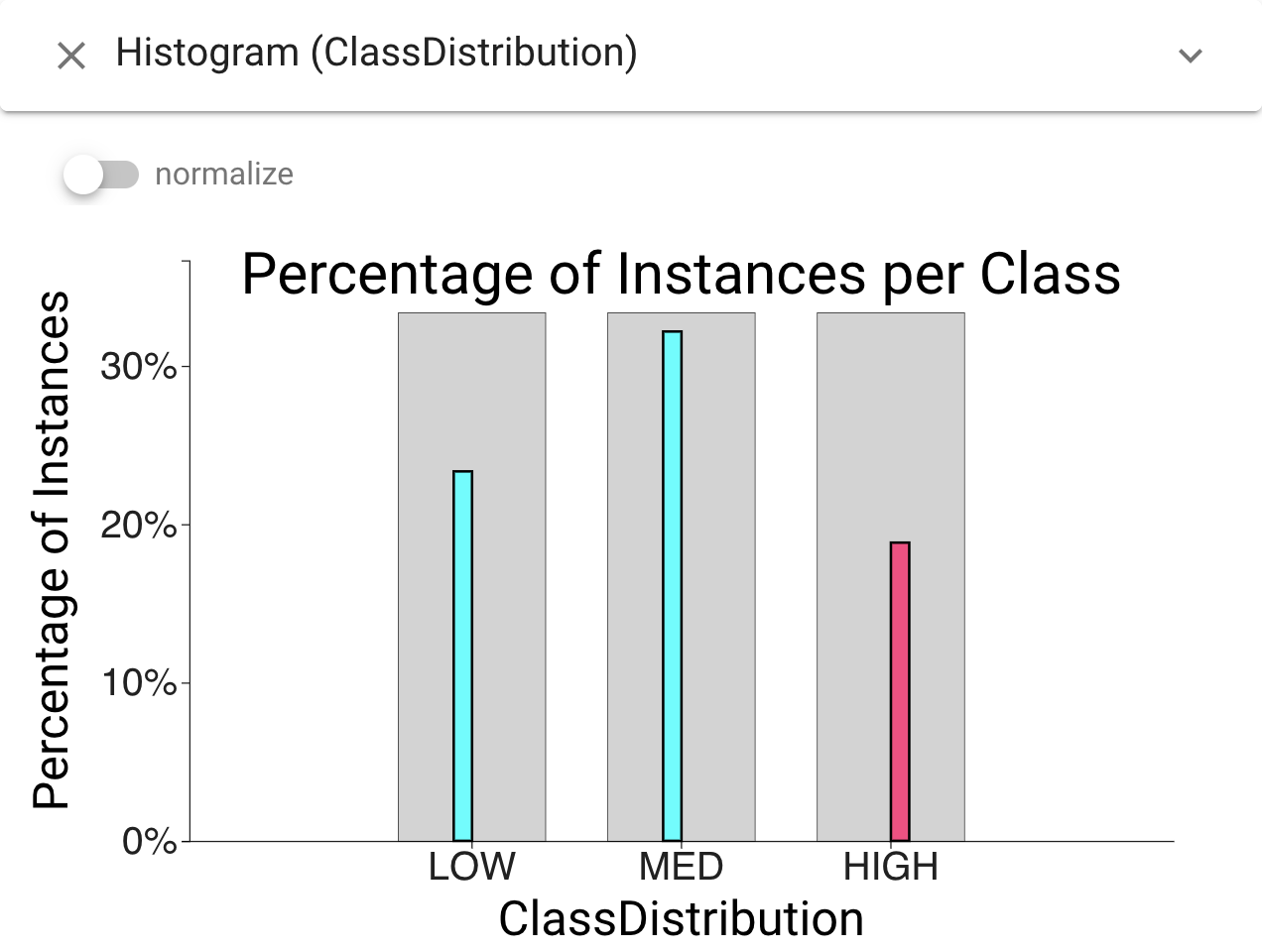}
			\includegraphics[width=\textwidth/5, valign=t]{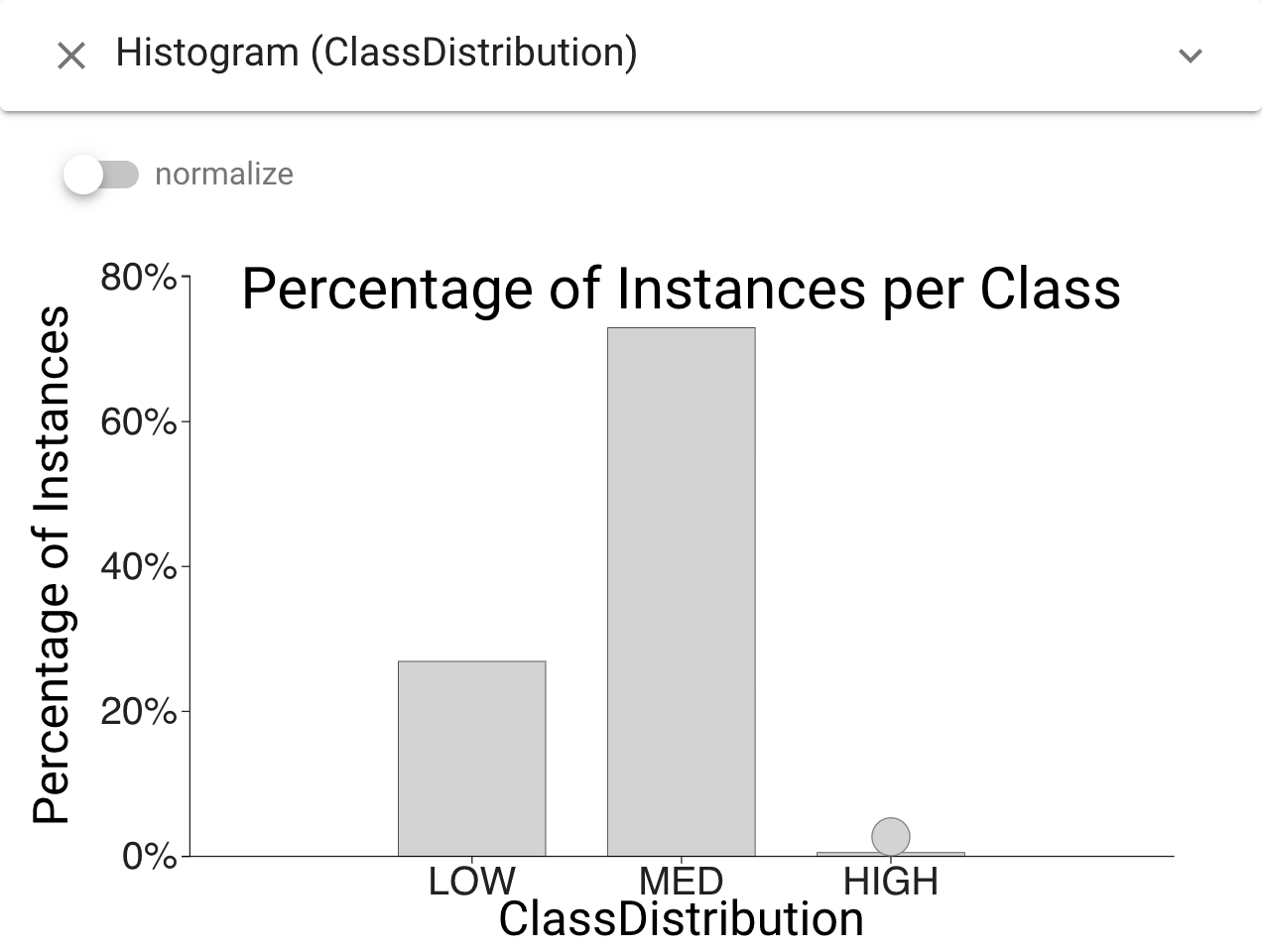}
			\includegraphics[width=\textwidth/5, valign=t]{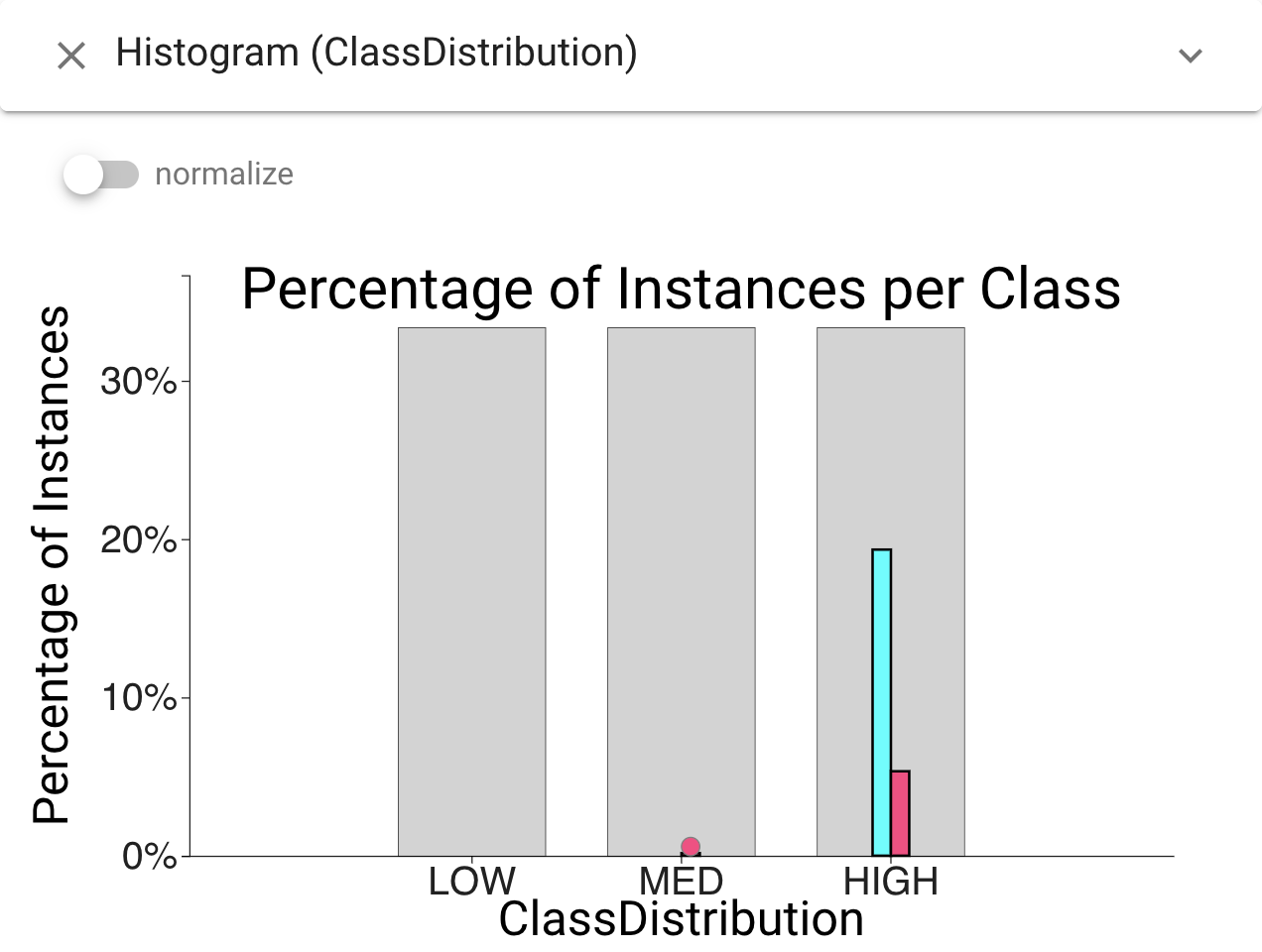}
		}
		\caption{\label{fig:imdb}
			Movie model selection use case (Section \ref{sec:imdb}). From left to right: (1) The Classifier Performance view shows low performance for all classifiers; (2) We then select easy (cyan) and hard (magenta) instances in the Cumulative Accuracy view; (3) A Histogram view of the test set shows us that all hard (magenta) instances are in the \emph{high} class; (4) A Histogram view of the training set shows class skew; (5) A Histogram view shows that the errors of the new classifier (magenta) are still biased, but less than the best prior classifier (cyan).
		}
	\end{figure*}
}

\newcommand\recidfig{
\begin{figure*}[t]
		\centerline{
			\includegraphics[width=\textwidth]{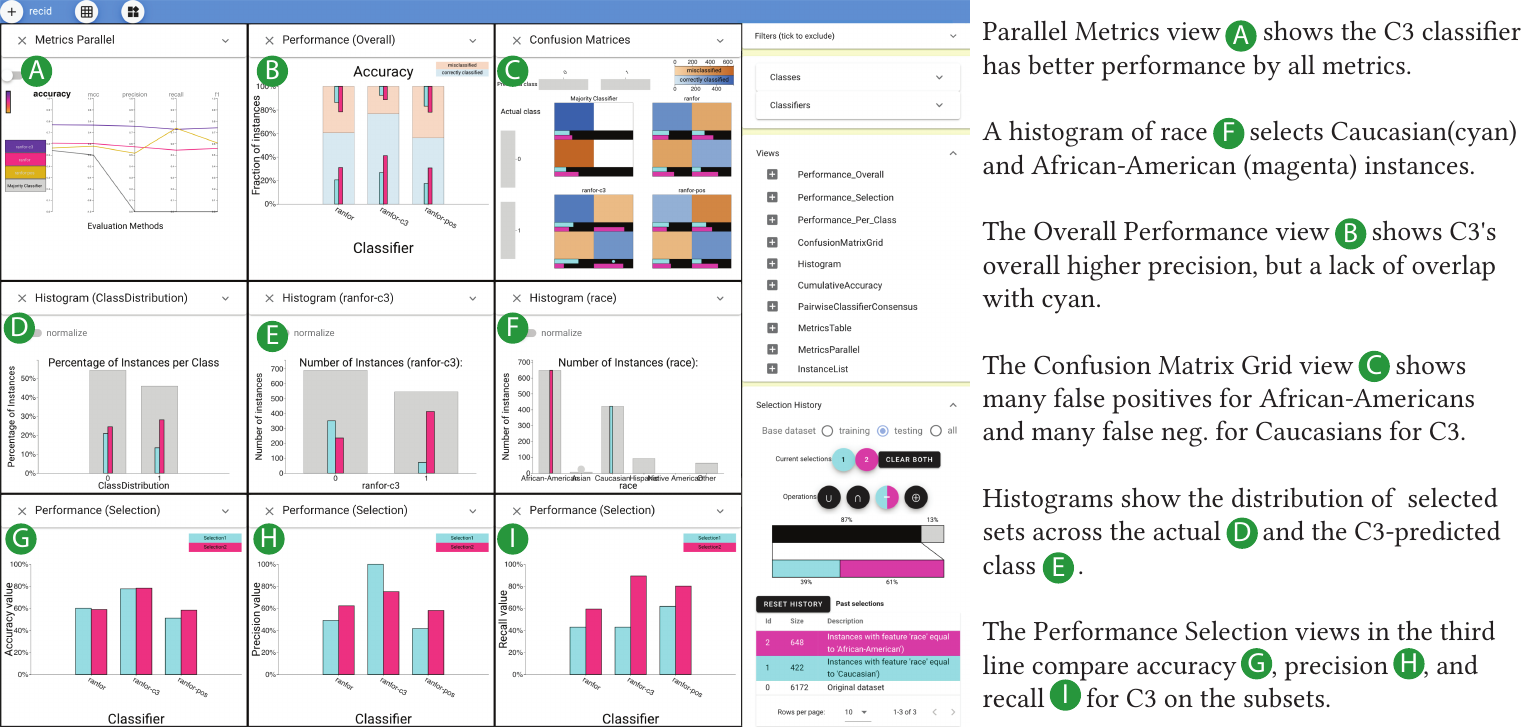}
		}
		\caption{\label{fig:recid}
				\sysname in the recidivism use case (Section~\ref{sec:recidivism}).
		}
\end{figure*}
}

\newcommand\figuredatehistogram{
\begin{figure}[t]
	\includegraphics[width=\columnwidth/2]{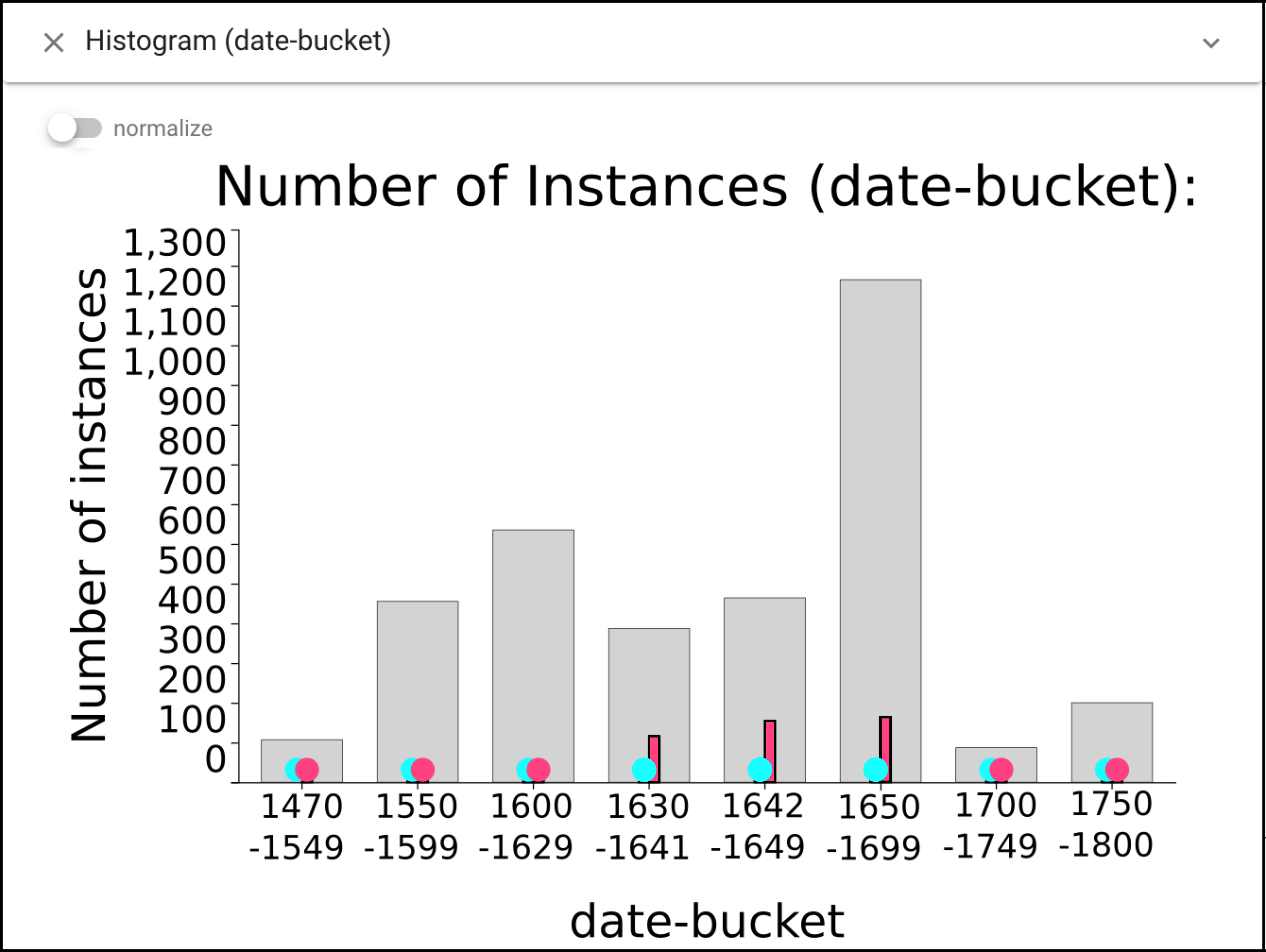}
	\includegraphics[width=\columnwidth/2]{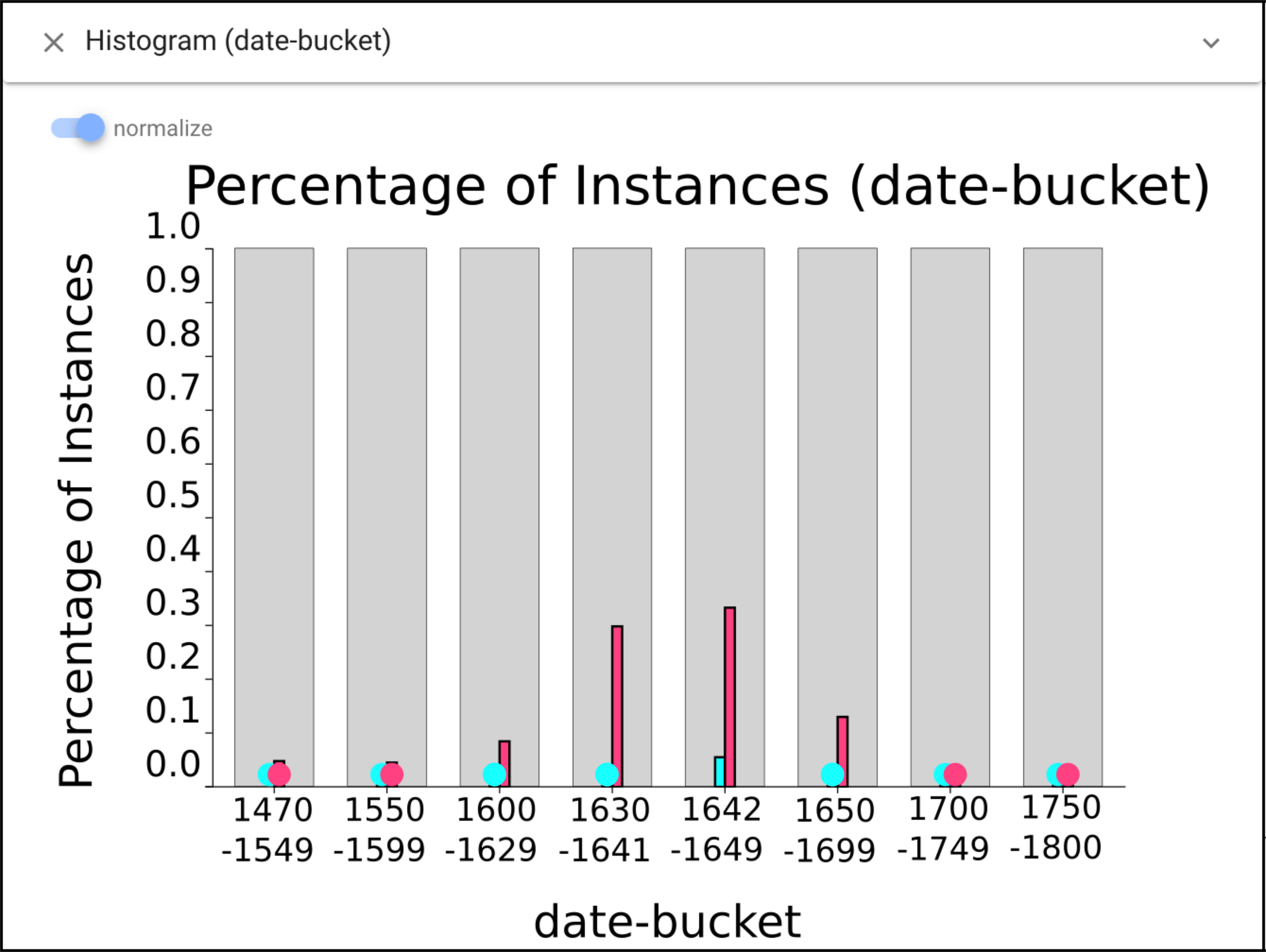}
\caption{\label{fig:date-histogram} Errors on all (magenta) and short documents (cyan) across temporal bins of documents.
				The right view is a normalized version of the left one.
				We can see that around 1642 there is a higher error rate in general, but the error rate on short documents is low. }
\end{figure}
}

\newcommand\figuretcpdate{
\begin{figure*}
	\includegraphics[width=\textwidth]{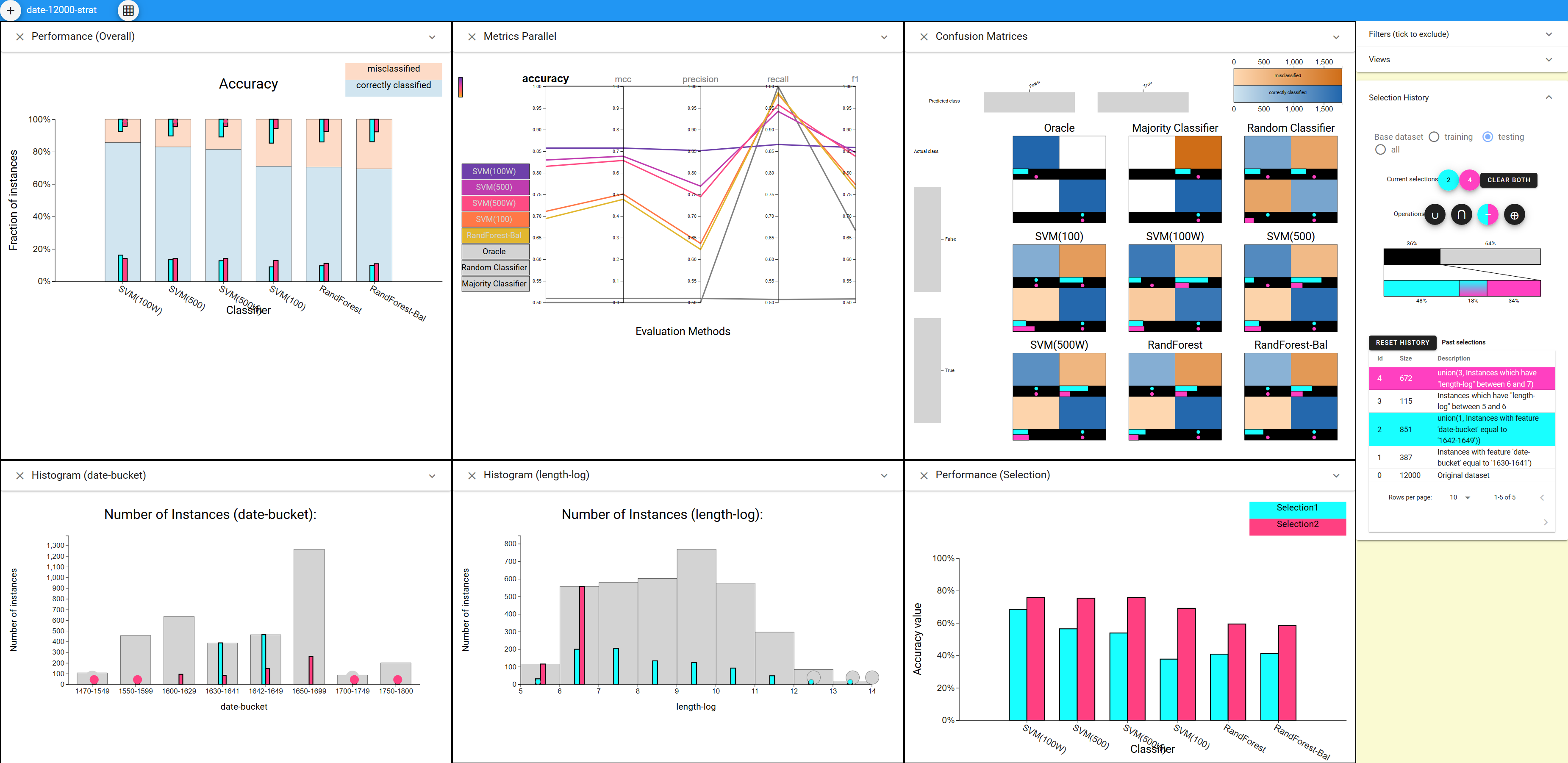}
	\caption{\label{fig:date}
			\sysname in the literature use case (Section~\ref{sec:dates}).
			In the Classifier Performance view (top left), we see that \emph{SVM(100W)} performs best in terms of accuracy.
			(1) From the Metrics Parallel and the Confusion Matrix views we learn that it is also the most balanced in terms of other metrics.
			(2) To assess whether classification is biased around the decision date (1642), we select the errors of the model (magenta), and view them on a date Histogram view and see that most errors happen in that period.
			(3) To test whether document length plays a role for these errors, we then select the shortest documents in the set in a document length Histogram view.
			(4) In the Performance Selection view (bottom right) reveals that shorter document (magenta) seem to be easier then longer ones.
		 	}
\end{figure*}
}

\section{Introduction}\label{sec:intro}

\maketitle

Machine learning practitioners often perform experiments that compare classification results.
Users gather the results of different classifiers or data perturbations on a collection of testing examples.
Results are stored and analyzed for tasks such as model selection, hyper-parameter tuning, data quality assessment, fairness testing, and gaining insight about the underlying data.
Classifier comparison experiments are typically evaluated by summary statistics of model performance, such as accuracy, F1, and related metrics.
These aggregate measures provide for a quick summary, but not detailed examination.
Examining performance on different subsets of data can provide insights into the models (e.g., to understand performance for future improvement), the data (e.g., to understand data quality issues) or the underlying phenomena (e.g., to identify potential causal relationships).
Relying on aggregated data can miss important aspects of classifier performance.
For closer examination, practitioners rely on scripting in their standard workflows.
The lack of specific tooling makes the process laborious and comparisons challenging, limiting how often experiments are examined in detail.

This paper presents \emph{Boxer} (\autoref{fig:teaser}), a system for the detailed examination of classifier comparison experiments.
Our approach allows a user to explore a collection of classifier results to identify interesting subsets of the data and compare performance across them.
Our system enhances standard views with interactions for selection and comparison.
The design provides a uniform mechanism for identifying subsets, choosing appropriate metrics, and assessing performance over different parts of the data.
Users combine views and build selections to pose comparisons for visual assessment.

Our work applies to the results of classifier experiments: the system operates on pairs of classifier inputs (testing and training data) and outputs (the predictions made).
Boxer treats classifiers as black boxes: it does not consider the internals of how the classifiers work, allowing it to be applied broadly and compliment existing, specialized tools.
To show the potentially large data, the system provides a set of summary views.
These views are variants of familiar displays, such as bar charts and confusion matrices.

Boxer is built around the unifying abstraction of a \emph{box}.
Boxes are a ``container'' for a subset of data elements, and a mark~\cite{Bertin2010, GrammarOfGraphics} that shows information about this subset.
They thus couple a data abstraction with a graphical abstraction.
A visual element, such as the bar of a bar chart or the square of a confusion matrix (Figure 2), connects input (select subset), display (show subset size), and comparison (relate subset to others).
Users create \emph{selections} of data elements of interest by selecting such boxes.
The selections are displayed relative to all boxes across the interface: users can see the overlap between the section and other boxes.
Boxer supports dual active selections, allowing for quick comparison and building more complex selections by combining simpler ones using binary set operators (e.g., intersection, union).
For example, in Figure 1, the user has selected the top parts of the leftmost bars in the Accuracy view (A) as the first (cyan) and second (magenta) selections.
These boxes represent the instances classified incorrectly by the two classifiers.
Cyan and magenta glyphs throughout the interface show the overlap of these selections with other boxes.
These selections can be combined, for example by intersecting their contents to identify instances classified incorrectly by both classifiers.

In summary, the three main innovations of Boxer are:
(1) an approach to classifier comparison that combines subset identification, metric selection, and comparative visualization to enable detailed comparison;
(2) an architecture of multiple selections and set algebra that allows users to flexibly link views and specify data subsets of interest;
and (3) interaction techniques and visual designs that make the approach practical.
These ideas should be applicable in other systems for interactive comparison.

\section{Related Work}\label{sec:relwork}
Interactive tools for understanding machine learning models are motivated by many reasons, see Gleicher~\cite{Gleicher2016} or Lipton~\cite{Lipton2016} for surveys.
The range of needs is addressed by approaches that can be roughly categorized into four groups:
\begin{enumerate}
\item Transparent models use learning representations designed for easy examination, such as generalized additive models~\cite{Lou2013,Caruana2015}, rule-based learning~\cite{Wang2015-fallingRuleLists,Obermann2016}, RETAIN~\cite{Choi2016-retain}, approximating decision trees~\cite{Martens2007,Chen1968,Craven1996thesis,Craven96-nips,Ming2019}, and sparse linear models~\cite{Gleicher2013}.

\item Per-instance explanations provide methods that explain individual decisions or local groups, examples include expert system tracing~\cite{Weiner1980, Wallis1984}, variable sensitivities~\cite{Tamagnini2017}, influential feature identification~\cite{Selvaraju2016a, Arras2017}, locally linear models~\cite{Ribeiro2016}, and instance examination~\cite{Krause2017}.

\item Internal inspection approaches provide tools to examine specific types of complex models such as neural nets~\cite{Strobelt1018, Liu2017, Ming2017}.
Variants of these approaches exploit knowledge of model structure, such as the Treepod~\cite{Muhlbacher2018} system that helps users understand trade-offs in decision tree classifiers.

\item Black box methods which do not consider the internals of the model, but instead rely on observations of their input/output pairs.
\emph{Our work falls into this last category, providing a general approach that works with a range of model types}.
\end{enumerate}

Spinner et al.~\cite{Spinner2020} develop a pipeline for model analysis and comparison that organizes such methods into three categories that focus on model input, model output, and model internals.
In the latter category, a range of methods have been proposed to ``open up'' those black boxes and provide a view of the inner workings of ML models.
Examples include representations of prototypical data instances~\cite{Mahendran2016, Yosinski2015, Alain2016}, structural overviews of complex models~\cite{MLiu2017a, Wongsuphasawat2017}, and explaining the influence of model structure on its output~\cite{Hohman2020, Rauber2017, Zeiler2014, Aubry2015}.
DeepCompare~\cite{Murugesan2019} allows instance-level comparisons of deep learning models and connecting decisions back to structural properties of the models.
\emph{In contrast, our work is strictly black box and focuses on using the results of existing experiments on models, and making comparisons between multiple models.}

Black box methods are often designed for very specific goals.
Ma et al.~\cite{Ma2020}, for example, support the identification of feature combinations that elicit a specific response from a model.
Ye et al.~\cite{Ye2019} enable users to assess and increase the quality of training data labels.
Another important goal is the analysis of fairness of predictions.
Friedler et al.~\cite{Friedler2019} review classifier fairness and provide measures.
Cabrera et al.~\cite{Cabrera2019} present a method for analyzing fairness by generating subsets of the data with different prediction performance.
In contrast, Ahn and Lin~\cite{Ahn2020} focus on identifying instance-level bias.
\emph{In contrast, our approach supports a broad range of tasks with flexible mechanisms to analyze different aspects of data and classifier performance.}

Some prior black box approaches support a range of tasks.
ModelTracker~\cite{Amershi2015} is an interface designed to provide an overview of model performance and detailed inspection of instances and their features through interaction.
\emph{In contrast, we provide a range of views that are linked through subsets of the classification data and allow in-depth comparisons between models.}
The What-If-Tool~\cite{Wexler2020} enables users to compose a range of visualizations, including bar chars and confusion matrices on subsets of their data by slicing sets based on feature values.
\emph{In contrast, we enable the creation of more complex subsets based on set algebra, and enable comparisons between models based on these sets.}
Manifold~\cite{Zhang2019} provides specialized variants of standard displays to show correlations between model decisions and their relation to the underlying data.
\emph{In contrast, we provide more flexible composition of views with comparative elements and selection construction.}

Black box methods are also used to test the stability and robustness of a trained model~\cite{Breiman2001}.
Such sensitivity analysis methods perturb the values of input features.
They can identify relationships between outputs and features~\cite{Henelius2014, Strumbelj2014}, even in complex models~\cite{Shrikumar2017, Olden2002, Sundararajan2017}.
Lee et al.~\cite{Lee2019} provide a systematic approach to test for the contributions of single features to model errors, and for potential interactions between features.
A conceptually similar approach is generating prototypical inputs for a model~\cite{Simonyan2013}.
\emph{Our approach is designed to examine the results of such experiments.}

There is a long standing effort in ML to develop metrics for assessing model performance~\cite{Powers2011, Witten2011}.
Some metrics address specific tasks such as interpreting the performance changes~\cite{Steyerberg2010}.
Similarly, visualization approaches extend basic chart types (e.g., ROC curves and confusion matrices) to present more information and serve as interactive tools for achieving specific user tasks.
For example, tools exist for
characterizing errors in regression models~\cite{muhlbacher2013partition},
examining classification results within their original context~\cite{Sarikaya2014},
adjusting weights from confusion matrices~\cite{Kapoor2010, Talbot2009},
and showing probability distributions in multi-class classifiers~\cite{Alsallakh2014-visPobClassifiers}.
In particular, Squares~\cite{Ren2017} enables in-depth comparison of classification models based on visual summaries of instance-level predictions and classifier probabilities.
However, the approach does not support subsetting of the dataset and comparisons between subsets.
\emph{In contrast, our work focuses on comparing models by identifying the subsets of instances that make up these metrics.}

Visual comparison approaches have been developed for many model types.
For example, approaches exist for specialized types of models, such as sequence models~\cite{Ming2020, Strobelt2019}, or for specific data types, such as set data~\cite{Alsallakh2014}.
Other approaches focus on clustering \cite{Kwon2017,Cavallo2019}, topic models~\cite{Alexander2016-compare}, word vector embeddings~\cite{Heimerl2018}, and climate models~\cite{Dasgupta2020}.
\emph{In contrast, our work focuses on making comparisons between discrete choice classifiers from existing classifier experiment data.}

\section{Comparison of Classification Models}\label{sec:abstraction}
We consider comparing the results of classifier experiments.
A \emph{classifier} is a machine learning model that predicts a \emph{label} from a data \emph{instance}.
Supervised machine learning builds classifiers using a \emph{training set} of instances for which the labels are known.
Classifiers may be evaluated by assessing their predictions on this training set (known as \emph{resubstitution}), but are more commonly evaluated on a \emph{testing} or \emph{hold-out} set of instances.
A \emph{classifier experiment} assesses the performance of one or more classifiers over a set of instances.
An experiment may involve multiple classifiers, comparing  performance on a common testing set, or involve the same classifier with multiple variants of the testing set.

The results of a classifier experiment are a set of data instances, a correct label for each instance, and, for each classifier, a prediction.
Boxer uses this data as input.
The data instances themselves are useful, as many tasks involve building connections back to the underlying data.
However, the measure in most classifier experiments is correctness: whether the predictions match the correct labels.
These results are summarized by counting the different outcomes across a set of answers.
A \emph{confusion matrix} counts all possible outcomes (if there are $l$ labels, there are $l \times l$ possible outcomes to be counted).
Metrics of classifier performance reduce this matrix to a single number.
For example, \emph{accuracy} counts the number of correct predictions (the prediction matches the correct label), and divides by the total number of instances.
Other metrics, such as F1, precision, recall, error rate, and mathews coefficient, summarize the counts of confusion matrices in different ways.
See Chapter 5 of Witten, et al. \cite{Witten2011} or Powers \cite{Powers2011} for comprehensive discussions.
Even more metrics are possible if additional information, such as confidence, is available for the predictions.
Selecting an appropriate metric is an important part of assessing classifiers.

\subsection{Tasks of Classifier Comparison}

Many tasks in classifier development and assessment involve comparison of classifier results.
Model selection, tuning, and fairness assessment rely on examination of testing results between classifiers.
Even the selection of an appropriate metric can be informed empirically.
Other tasks, such as data quality assessment, feature engineering, and gaining insights from data can make use of a collection of classifiers and their results.

However, classifier comparison typically considers summary metrics over the entire testing set.
This hides information about which instances different classifiers get right or wrong.
This information may be useful across a range of tasks.
For example:
\begin{itemize}
\item In \emph{model selection,} having information beyond summary metrics can determine performance in more important and relevant cases. This can help adapt test metrics to predict real world performance.
\item In \emph{model tuning,} understanding where a model works (or doesn't) can help choose strategies for improving performance and confirm that interventions work as expected.
\item In \emph{fairness assessment,} different subgroups of instances can be compared to identify ones being treated unfairly.
\item In \emph{data quality assessment,} identifying interesting instances can point to problems or opportunities in the underlying data. For example, view can show that errors as associated with missing data.
\item In \emph{studying the underlying data,} identifying specific instances or differences between groups can help test theories or identify underlying mechanisms behind the data.
\end{itemize}
We will provide examples of such scenarios in Section~\ref{sec:use_cases}.

The range of tasks that can use subset selection and assessment is broad and diverse.
However, in working with practitioners performing these tasks, a core pattern of elemental operations emerge.
Users must (1) identify an appropriate subset of the instances; (2) select an appropriate metric; (3) compare performance across classifiers; and (4) relate these results in both the broader context as well as specific details.
This process is often exploratory and iterative: a user examines a metric across a number of subsets, making comparisons that suggest different combinations.
The typical practice of using scripting to query data to create subsets and perform evaluations does not support rapid exploration or visualization that enhances comparison or helps contextualize results.

Rather than designing specific support for the broad range of tasks, our strategy is to provide flexible support for elemental operations common across tasks, and provide mechanisms to combine these operations to support workflows that address tasks.

\subsection{Boxes and Selections}\label{sec:selections}
Many classifier comparison explorations involve two basic operations: (1) defining subsets of instances, and (2) comparing performance and specific predictions between these subsets.
By composing these two elements, we can construct many more complex tasks.
For example, by comparing the subset of instances a classifier gets wrong with the subset formed by histogram bins (e.g., of a feature), we can see where in the dataset an error occurs.
Our design focuses on providing fluent and flexible support for these basic operations.

We support subsets through the abstractions of \emph{boxes} and \emph{selections}. A \emph{box} connects a data abstraction to a visual abstraction. A box represents a specific subset of the instances, which can be defined as a query over the entire collection. A box also connects to a visual element that summarizes the subset. For example, a bar of a bar chart represents a subset of instances (the query is the x range of the bar) {\emph{and} a visual element (the rectangle encoding the count of the subset with height). A \emph{selection} is a subset of instances of interest to the user. At any given time, there may be an \emph{active} selection that is being explicitly highlighted. 
\differentboxes

Boxes and active selections connect: boxes serve to create selections (e.g., clicking on a box sets the active selection), and the active selection can be presented visually in each box to enable comparison by showing the intersection between the box and the selection. This is possible because each box represents a subset of the data. The queries serve as textual descriptions that describe selections, for example in history views.
Our system supports two active selections. Both selections are shown in all boxes (Figure~\ref{fig:different_boxes}). The left and right mouse button are used to make corresponding selections.
Consistent coloring is used throughout the interface: cyan for the first active selection, magenta for the second. We use vivid colors that contrast other interface elements even at small sizes \cite{szafirModelingColorDifference2018}. 
Multiple selections extend prior abstractions of aggregation (e.g., \cite{Slingsby2009,Elmqvist2010}) to better support composition and comparison. 

Dual active selections are a key feature of Boxer. A single active selection does not allow for comparisons between selections, and does not allow for an interface for selection composition. More than two active selections may enable richer comparisons, but comes at the expense of visual clutter, the need to reserve more colors (precluding their use elsewhere in the system), and a need to find new methods for showing and interacting with the sections. 
Dual selections are sufficient for creating complex subsets by composing binary set operations (e.g., intersection, union, subtraction).

Information about the active selections are shown in the Selection Control view (Figure~\ref{fig:selection_control}). This view provides a textual description of current and previous selections, a history of previous selections allowing them to be recalled by clicking, and a widget that shows the relationship between the two active selections. The Selection Control view allows for set arithmetic operations to be performed between the two selections using the relationship widget.
For example clicking on the overlap of the two selections creates a new selection that is their intersection.
\selectioncontrolfigure

The box abstraction helps create a uniform and flexible mechanism for expressing set comparison as well as a consistent user experience.
Boxes in our system support four features. First, they have an associated \emph{query} that finds the instances in their subset and allows for textual description. Second, they have some \emph{visual representation} that displays the size of the set.
Third, the visual elements of a box are \emph{clickable,} serving as input to identify that the subset associated with the box is of interest to the user. Fourth, each box displays a \emph{comparison} with the active selections. 
Our system supports two active selections, so each box must show three quantities: the number of instances in the box's associated subset, the size of the intersection of the box's set with the first selection, and the size of the intersection of the box's set with the second selection. Examples are shown in Figure~\ref{fig:different_boxes}.

Each rectangular region of a bar chart, stacked bar chart, or confusion matrix is a box. Other visualizations, including pie charts, hexbin plots, and treemaps, could similarly provide a set of regions, each corresponding to a box. The four features of boxes give an interaction mechanism and visual grammar for displaying comparisons. All visualizations built from boxes serve as mechanisms to specify selections, and to make comparisons between the set defined by the box and other sets of interest to the user. 

Users build selections by selecting boxes and combining them with binary set operations to make more complex selections. Users make detailed comparisons by seeing how the selections compare to each other or against boxes across different views.

\section{Boxer Views}\label{sec:boxer_system}
Boxer provides users with a number of view types that can be combined in the workspace as needed.
The box mechanism allows for specifying sets of interest and performing comparisons in these familiar views, allowing users to couple simple displays to address complex tasks.
Our design explores the use of a few basic views and a general composition mechanism, rather than designing more specialized views.
Other view types could be added in the future. However, introducing new views has a cost: users must be able know when to apply them.
Boxer had other views that have not proven useful in practice, and were removed. 

\subsection{Bar Chart-based Views}
Many of \sysname's main views use bar charts that divide the instances in different ways. Bars correspond to boxes: they can be used to select their contents as well as to display their overlap with the two active selections, as shown in Figure~\ref{fig:different_boxes}. Boxer's bar charts illustrate non-zero values with circles (e.g., Figure \ref{fig:imdb}.2, \ref{fig:imdb}.4, \ref{fig:date-histogram}) which are hidden if the bar becomes large enough. This allows small subsets (such as rare instances) to be seen and selected as the data grows. Charts allow for normalization to enable comparisons between sub-bars (Figure~\ref{fig:date-histogram}).

The \emph{Classifier Performance} view (Figure~\ref{fig:tcp-feat}a, \ref{fig:imdb}.1, \ref{fig:fuzz} left) shows the performance of each classifier in a stacked bar.
A variety of metrics can be chosen. For metrics that are ratios of subset counts (e.g., accuracy, precision, and recall), the bars are stacks of boxes. This view provides a simple overview of classifier performance, and an easy way to select sets of instances (e.g., what classifier predicts correctly). The view allows sorting by value to facilitate identification and comparison of the best or worst classifiers.

The \emph{Histogram} (Figure \ref{fig:tcp-feat}e, \ref{fig:tcp-feat}f, \ref{fig:imdb}.2-5) view shows the distribution of the data across a feature.
This includes the data features as well as the actual and predicted classes.
The user can place multiple histograms to show different distributions. Continuous features are bucketed, and discrete features can be sorted in various ways (e.g., sort by quantity to emphasize the largest categories). Histograms provide an important mechanism for selection as well as comparative display. Combining bars from multiple histograms using set operations can specify complex selections.

The \emph{Cumulative Accuracy} view (Figure~\ref{fig:tcp-feat}d, \ref{fig:imdb}.2), shows how many classifiers correctly labeled each data instance. For example, the viewer can identify items that no classifiers predicted correctly or that all classifiers predicted correctly. This view may be used to select challenging instances, or to see if a selected set contains easy items. The Cumulative Accuracy view includes a pareto line allowing the user to quickly assess and select the cumulative sum.

The \emph{Selection Performance} view (Figure \ref{fig:recid}g, h, i, and \ref{fig:fuzz} right) shows the performance of each classifier across both selections. The user can select a variety of different metrics (accuracy, F1, etc.). Because the bars are associated with selections (and colored accordingly), they do not serve as boxes. In contrast, the \emph{Per-Class Performance} view provides boxes while showing the performance of each classifier for each of the actual classes of the instances. The user selects a metric (accuracy, F1, etc.) for this faceted bar chart.

\subsection{Matrix-based Views}
Matrix-based views arrange boxes in a fixed grid. Because the size is fixed, color encodes for the size of the subset the box corresponds to. The color encoding allows for rough comparison between squares: detailed values can be revealed by hovering. Small bars, color-coded to match the selection scheme, show the overlap between the active selections and the box's subset -- a full bar means that all of the box's instances are in the selection. Matrix cells can be clicked for selection.

The \emph{Confusion Matrix Grid} view (Figure \ref{fig:tcp-feat}c) provides the standard view of classification results for each classifier.
It shows each classifier's performance, broken down by label.
This allows us to compare classifiers based on their prediction profile per class.

The \emph{Pairwise Consensus} view shows the agreement and disagreement between each pair of classifiers as a matrix.
It conveys the number of instances for which two classifiers predict the same label. The matrix is split on the diagonal to distinguish agreement on correct vs. incorrect instances.
This view is useful in identifying correlations between classifiers, for example to assess ensembling.

\subsection{Other Views}
Boxer provides views designed to help with metric selection. The \emph{Standard Metrics} view provides a simple table of many metrics across all classifiers. The \emph{Parallel Metrics} view (Figure~\ref{fig:tcp-feat}b) presents the same information in a parallel coordinates chart. Similar to \cite{Dasgupta2020}, parallel coordinates help identify correlation between metrics. The view uses an ordered coloring based on a selected metric to aid in order comparison.

The \emph{Instance List} view shows a tabular display of the instances in the active selections.
Instances are color-coded to indicate which selections they are part of.
Users can select instances, allowing for fine-grained, instance-level modification of selections.

The \emph{Selection Controls} panel was described in Section~\ref{sec:selections} and Figure~\ref{fig:selection_control}.
In addition to selection display and interaction, it allows users to switch between training set, selection set, or both.
The \sysname interface allows viewing all data instances, or limiting the views to either training or testing data.

\section{Use Cases}\label{sec:use_cases}
This section provides some example scenarios where Boxer can be used to address tasks in classifier development.
These examples are chosen to highlight how Boxer's key ideas can be applied in common machine learning settings.
The examples come from student data scientists working with standard data sets, as well as from our collaboration with literature scholars to use statistical techniques to help analyze historical text corpora.

All examples were completed using the Boxer system.
Boxer is built as a web application, written in TypeScript and using the Vue.js application framework and D3.js for drawing visualizations.
The system loads all data, both classification results as well as feature and meta-data for all data instances into memory at startup and performs all computations within the browser.
Boxer is efficient enough to handle data sets with tens of thousands of instances. We discuss scalability issues in Section~\ref{sec:scalability}.

\subsection{Model Selection and Tuning: Movies} \label{sec:imdb}
\imdbfigure
We consider a simple use case that shows how Boxer's views can be combined to diagnose classifier problems. 
We use boxer in developing a classifier to predict movie ratings from IMDb.
Our data consists of 5,044 movies with 27 features.
25\% are sequestered for final assessment.
Classifiers predict a movie's rating (low, medium, high).
A stratified sampling of 200 movies per class is held out from the 75\% remaining.
A variety of classifiers were constructed, none with acceptable accuracy.

Using boxer, we examine the results (Figure ~\ref{fig:imdb}).
The Classifier Performance view confirms the poor performance.
The Cumulative Accuracy shows large numbers of instances that are easy and hard (all or no classifiers are correct).
We select these easy and hard subsets.
In a Histogram view, we can see the hard elements are in the \emph{high} class.
Examination of the training set shows that there are few examples of this class.
We build a new classifier (offline) that accounts for this skew.
Using boxer, we can confirm that this has superior performance, although its errors are still biased.
While conventional tools can show skew, the example shows how Boxer's flexible mechanisms allow performance effects to be connected to data issues. 


\subsection{Model Selection and Data Discovery: Literary Features} \label{sec:tcp}

This use case considers a corpus of 59,989 documents from a historical literary collection: Text Creation Partnership (TCP) transcriptions of the Early English Books Online (EEBO).
Of these documents, 1,065 have been identified as \emph{plays},  1,974 as \emph{science} documents, and most are \emph{neither}.
The data counts the 500 most common English words in each document.
While all documents have been classified by experts, we construct classifiers using the data to support theories that different types of documents use words in different ways~\cite{Witmore2010,Gleicher2013}.
Specifically, we are interested if a small set of words can  identify document classes.
Boxer allows us to compare the performances of classifiers built from different sets of words to confirm the impact of word choice on performance.

We create decision tree classifiers using a variety of univariate feature selection strategies.
Each selects 10 words to count for features.
The feature selection methods were: most relevant by a CHI-squared univariate feature selector (C), most common features (N), randomly chosen features (R), and, as a baseline, the features deemed worst (out of the 500 candidate words) by the CHI-squared test (W).
A testing set of 200 documents per class was used.

We compare the results in Boxer (Figure~\ref{fig:tcp-feat}).
The Parallel Metrics view shows a consistent ordering of the classifiers across all metrics: C is slightly better than R and N, which are much better than W.
Using the Classifier Performance view to see the accuracy details, we can select the mistakes made by the top classifiers (cyan for C's mistakes, and magenta for R's mistakes).
We see that the errors are relatively evenly distributed among the classes in a Histogram view of the class distribution.
This is surprising given the skewed training distribution.
We also see in the Performance view that different classifiers make different errors (e.g., only half of C's errors are made by N).
Overall, the Cumulative Accuracy view shows that there are very few instances that all classifiers were wrong on (2.6\%), and a Histogram view of document lengths lets us see that performance is relatively consistent over the range of document lengths.

\subsection{Feature Sensitivity Testing: Plays} \label{sec:fuzz}
%
%
%
\fuzzfigure

We use \sysname to provide more insight into the results of a variable sensitivity experiment.
The data set is a collection of 554 plays written in the Early Modern Period (1470-1660).
Five linguistic features (selected form~\cite{Ishizaki2011}) are used.
We classify plays with one of four genres (Comedy, History, Tragedy and Tragi-comedy).
Ground truth is known; the goal is to use the classifier to determine relationships between the linguistic features and genre~\cite{Witmore2010, Gleicher2013}.
Our experiment uses a Support Vector Machine (SVM) classifier trained with class weights to counteract a skewed training distribution.
A stratified sample of 20\% was removed as the test set.
The training set has very few of the under-represented classes.

After training, a feature sensitivity experiment identifies which features contribute to the classifier's performance.
We create a variant of the data set for each feature.
In each data variant, small perturbations (positive and negative) are added to its corresponding feature's value for all entries in the data set.
The resulting data sets have twice as many items as the original (one for positive additions to the feature, one for negative ones).
Each data variant is run through the classifier.
Here, we consider only the testing set.
Such experiments are preferred to simply examining model coefficients because they test the effects of the variables near the actual data.

The standard approach to analyzing such an experiment is to compute summary statistics over each feature's data variant and compare these to the baseline.
More advanced approaches~\cite{Lee2019} can check for the statistical significance of these differences.
The experiment results show that for feature \emph{Negavity} (N), the classifier performs much worse than the baseline using standard metrics (accuracy and Mathews correlation).
Two features \emph{PersonProperty} (PP) and \emph{DirectAddress} (DA) achieve similar performance to the baseline.
Standard procedure would conclude that the classifier is sensitive to N but not PP and DA.
However, closer examination in Boxer reveals otherwise. 
The Classifier Performance view (\autoref{fig:fuzz}) shows that PP and DA have similar accuracy to the baseline.
Selecting the correct and incorrect subsets for the baseline allows us to compare with the perturbed results.
For DA, the overlaps are substantial, for PP there is less overlap.
While PP gets the same number of instances correct, it is correct on different ones, suggesting the model is sensitive to this feature.

\subsection{Model Selection: Mushroom Imputation} \label{sec:dqa}




We consider adapting a classifier to identify poisonous mushrooms.
The initial training of the baseline classifier uses the \emph{color} feature, but we would like to build a new classifier that does not use this feature.
We consider two approaches to imputing the missing feature: using the mode of the data and training a decision tree to classify the color based on other features.
In addition to the baseline classifier, we build two new models: \emph{mode} and \emph{smart} that use the imputed versions of the color feature.
Note that mode is effectively not using the color feature, as it is constant across its data set.

We assess these strategies with a testing set of 2,000 (of 8,124) randomly selected instances.
The Parallel Metrics view shows that the imputed models perform worse than baseline on all metrics.
The smart model performs better than mode on all metrics except recall, which is likely to be important (we don't want to eat a poisonous mushroom).
To understand these differences, we can select the instances where the baseline and smart classifiers are incorrect and see that the latter is almost a proper subset of the former.

We wish to understand if the lower performance of the smart classifier can be attributed to imputation mistakes.
We select the instances where the baseline is correct and where smart is wrong, and intersect them.
Examining these sets in a Histogram view of the color feature, we see that most of the errors are white mushrooms.
Looking at the imputed feature over this set, we see that the smart imputer never labels these as white.
These mistakes of the imputer likely cause the misclassifications.

\subsection{Fairness Assessment: Recidivism} \label{sec:recidivism}
%
%
%

We consider a standard test case for fair learning: the Broward County recidivism dataset, popularized by ProPublica~\cite{Angwin2016}.
This data set was initially used to show unfairness of a commercial system, but has emerged as a benchmark for machine learning fairness~\cite{Friedler2019}.
The task is to predict whether a person will commit a crime within two years (two year recidivism).
The data set includes ground truth.
Classifiers built for this problem are often \emph{unfair} in that they skew errors towards racial and gender bias.
Specialized tools, such as FairVis~\cite{Cabrera2019}, are designed for assessing classifier fairness.
In this use case, we show how Boxer's flexible mechanisms can be used for similar purposes.

The data set contains 6,172 instances (chosen by the criteria of~\cite{Friedler2019}) and 14 numeric features (created by one-hot encoding the categorical features in the initial seven feature data set).
20\% are held for testing.
We consider three classifiers trained on the data, a \emph{baseline} random forest, and two hand-tuned variants (\emph{C3} and \emph{Pos}).
In the Parallel Metrics view, we can see that C3 achieves higher scores for all metrics.
We question whether it achieves these improvements in a fair manner.

\recidfig

Analysis of this case is shown in \autoref{fig:recid}.
We use a Histogram view of the \emph{race} feature to select Caucasian 
and African-American 
instances.
Various views in boxer clearly show the unfairness.
While a Selection Performance view shows similar accuracies for the selections, the precision and recall are very different.
C3 has high precision but low recall for Caucasians, and high recall but low precision for African-Americans.
That is, its errors are biased to predict \emph{no} for Caucasians and \emph{yes} for African-Americans.
While other classifiers make more errors, their errors are more uniformly distributed.
This can also be seen in the confusion matrices.

To explore further, we consider the effect of gender.
We select the subset of female instances by left clicking in a histogram and intersect this with the African American subset (using the relationship widget).
C3 has 0 recall on this subset of African American females, while other classifiers achieve more balanced performance.
While these findings could have been found using scripting~\cite{Friedler2019} or using specialized tools~\cite{Cabrera2019}, \sysname can identify them using subset selection and comparative visualization.

\subsection{Bias and Data Discovery: Literary Dating}\label{sec:dates}

In this use case, we again consider the TCP collection of historical documents (Section~\ref{sec:tcp}).
We construct classifiers that determine whether a document is written after 1642 based on the 500 most common words in the corpus.
While ground truth is known, effective classifiers can help understand how word usage changed at this critical date that marks the beginning of the English Civil War.
The collection is skewed (only 25\% of the documents were written before 1642).
For the experiment, we took a random sample of 12,000 documents, and held out 30\% using stratified sampling.
While the testing set is balanced (1,800 per class), the training set is highly skewed (only 15\% before 1642).
We constructed a number of classifiers using various methods.

\figuretcpdate
An image from an analysis session with \sysname is shown in Figure~\ref{fig:date}.
On the training data, several classifiers achieve nearly perfect performance.
However, the Parallel Metrics view (on the test set) shows that most classifiers provide high recall but low precision, suggesting that they were unable to successfully account for the class skew (in the training set).
The SVM100W (a support vector machine using class weights and more regularization) provides the best performance in all metrics other than recall.
The more balanced performance can also be seen in the Confusion Matrix grid.
We focus on this classifier for our assessment.

We would expect that performance may be biased near the class boundary, as documents written near the boundary year may be similar to those on the other side (unless there was a dramatic change at the boundary).
To check this effect, we select the errors of the classifier 
and view the selection in a histogram of dates (Figure~\ref{fig:date-histogram}). Normalizing this histogram (to account for the skewed distribution), confirms that most errors are in the buckets near the boundary.
In contrast, the skewed classifiers generally made many errors in other buckets before 1642.

One explanation for the errors may be document length: near the civil war, many short documents were written (e.g., legal decrees).
To explore this, we select the shortest documents
and create a subset of documents written near the boundary.
The skewed distributions in the lower left and center of Figure~\ref{fig:date} make the result hard to interpret, but normalizing these histograms show that short documents are over-represented in the time period.
However, intersecting the two sets (to select the short documents in the period) allows us to consider performance.
Intersecting this set with the errors (and comparing with the total errors) show that prediction performance is better on short documents (Figure~\ref{fig:date-histogram}).
This can also be seen in the Selection Performance view.
Alternatively, we can select errors and period documents to see that there is a skew towards short documents.
Using \sysname's ability to create subsets and compare them, we can examine details of classifier performance.
\figuredatehistogram

\section{Discussion}\label{sec:discussion}
We have presented a comprehensive approach for interactive comparison for machine learning classifier results, and a prototype implementation, \sysname.
The approach combines subset identification, metric selection, and comparative visualization to enable detailed comparison of classifier results.
We demonstrate the effectiveness of our approach for a range of datasets and model types through use cases.

\subsection{Scalability}
  \label{sec:scalability}
The classifier comparison problem can scale along all three ``axes of hardness''~\cite{Gleicher2018} with many instances, classifiers to compare, and complex relationships between classifiers.
Boxer has been used  to analyze experiments with up to tens of thousands of instances, a dozen classifiers, a dozen labels, and dozens of features.
Boxer's in-browser JavaScript implementation grows sluggish at these scales.
Providing interactive performance for larger data will require more efficient mechanisms to perform the set computations and will probably require sharing some of the computations in a back-end server.
The more interesting scalability challenges relate to how well the \sysname design can handle larger problems.

The visual summaries used by \sysname are independent of instance quantity: they show aggregate quantities (e.g., counts).
A bar chart appears the same if the Y axis represent 100 or a million.
However, one challenge is dynamic range: small values in big sets can be important (e.g., identifying a few errors in a massive training set).
Even at current scales, interesting bars may be a fraction of a pixel tall, and similar issues exist for color encodings.
To combat these dynamic range issues, box designs use special encodings for small values.
Zero (empty sets) are encoded differently than small sets, and small sets are given special encodings that provide them with sufficient area to be clickable.

Instance views, such as lists, represent a different scalability challenge.
With large numbers of instances, selections may contain many items.
Presently, \sysname handles the performance issues of long lists by on-demand paging, and the usability of long lists by allowing for sorting and filtering.
Future extensions could provide more automated assistance in finding interesting elements in long lists, such as representative subset sampling.
In general, scalability concerns have led us to avoid per-instance displays, like the scatterplots used in Manifold~\cite{Zhang2019} or the stacks in ModelTracker~\cite{Amershi2015}.

Scaling to large numbers of classes and classifiers represent design challenges.
The visual elements of bar charts and matrices must become smaller as the number of categories to show grows.
For moderate numbers, matrix re-ordering techniques can help our designs remain useful for larger data.
However, different visual designs may be required to afford comparisons between large numbers of features or classifiers.
At present, \sysname treats features independently: the user must select specific features to examine.
To better scale to large numbers of features, \sysname will need to incorporate interest operators to help guide users to interesting features.

\subsection{Other Problem Types}

\sysname is designed for the results of discrete choice classifiers.
Extending the prototype and concepts to other problem types is important future work.

\emph{Probabilistic Classification Results:} Boxer treats prediction probabilities as metadata features for analysis. This provides for limited analysis. Future extensions to handle probabilistic classifiers may adapt existing designs such as Squares~\cite{Ren2017} and Confusion Wheels~\cite{Alsallakh2014-visPobClassifiers}, as well as standard metrics, such as ROC curves, and and area under the curve (AUC/ROC).

\emph{Continuous Prediction Problems:} Presently, Boxer treats \emph{regression} problems by discretizing the output into a set of discrete classes. Existing visualizations for assessing regression models (e.g., \cite{muhlbacher2013partition}) could be extended with the box abstractions.

\emph{Problems without Ground Truth:} If ``correct'' labels are not known, Boxer selects a model as the ``gold-standard.'' This limits application in unsupervised tasks. 
Supporting ``differently correct'' answers within the \sysname framework will require designs for better exploring agreement, perhaps by adapting designs for clustering assessment (e.g., \cite{Kwon2017,Cavallo2019}).

\subsection{Limitations}

\sysname presently uses a number of basic visualization designs.
Other visualizations, such as scatterplots or non-linear dot plots~\cite{Rodrigues2018}, may be useful to better understand instance distributions.
Adapting visualizations to work within \sysname requires identifying mechanisms for specifying selections as well as displaying overlaps with active selections.
For example, with a scatterplot, overlap may be shown by coloring the dots and selection may be accomplished with brushing.
However, computing the overlap to display new selections must be made efficient, and concise textual descriptions of brushing results are required to integrate into the selection management mechanisms.

\sysname presently is a stand alone tool for viewing experiment results.
It provides no support for helping to design and run appropriate experiments.
Better coupling with the experimental process offers opportunities as the increased facilities to analyze experiments suggests the potential for non-standard experiments.
Similarly, \sysname is decoupled from the model building process.
Integrating Boxer into an automated learning pipeline (e.g., \cite{Cashman2019,Gil19,Visus,ATMSeer}) may provide a mechanism to more directly apply insights to improve models. 

A key factor in \sysname is usability.
The complexity of \sysname's interface is kept low by the choice of familiar visualizations and the use of a small set of basic abstractions that are used uniformly.
However, in order to make complex analyses, a user must combine these basic elements in potentially complex ways.
In principle, such complex assemblies are built up gradually from parts, but a user must know what combinations are possible and useful.
To address these issues we plan to include pre-configured layouts to answer specific questions (as in~\cite{Szafir2016textDNA}) and workflow-based guidance to suggest potential next steps to a user~\cite{Ceneda2017}.

\subsection{Conclusion}

Despite these limitations, the \sysname prototype shows the potential for the approach to help users with comparing classifiers in machine learning experiment results.
We are continuing to work with machine learning practitioners to refine the system and understand its potential.
The key innovations of \sysname, the consistent use of the box abstraction, the use of multiple selections with set algebra, and the specific interface designs for connecting boxes and selections should be applicable in creating scalable systems for exploring comparisons of objects beyond machine learning classifiers.

\paragraph*{Acknowledgements} This work was funded in part by NSF Award 1841349 and DARPA award  FA8750-17-2-010.

\bibliography{boxer_eurovis}

\newcommand{\etalchar}[1]{$^{#1}$}
\begin{thebibliography}{\uppercase{DWOB20}}

\bibitem[AB16]{Alain2016}
\textsc{Alain G., Bengio Y.}:
\newblock {Understanding intermediate layers using linear classifier probes}.
\newblock \emph{{arXiv preprint arXiv: 1610.01644}} (Oct 2016).

\bibitem[ACD{\etalchar{*}}15]{Amershi2015}
\textsc{Amershi S., Chickering M., Drucker S., Lee B., Simard P., Suh J.}:
\newblock Modeltracker: Redesigning performance analysis tools for machine
  learning.
\newblock In \emph{Proceedings of the Conference on Human Factors in Computing
  Systems (CHI 2015)} (April 2015).

\bibitem[AG16]{Alexander2016-compare}
\textsc{Alexander E., Gleicher M.}:
\newblock {Task-Driven Comparison of Topic Models}.
\newblock \emph{IEEE Transactions on Visualization and Computer Graphics 22}, 1
  (Jan 2016), 320--329.

\bibitem[AHH{\etalchar{*}}14]{Alsallakh2014-visPobClassifiers}
\textsc{Alsallakh B., Hanbury A., Hauser H., Miksch S., Rauber A.}:
\newblock {Visual Methods for Analyzing Probabilistic Classification Data}.
\newblock \emph{IEEE Transactions on Visualization and Computer Graphics 20},
  12 (Dec 2014), 1703--1712.

\bibitem[AHM{\etalchar{*}}17]{Arras2017}
\textsc{Arras L., Horn F., Montavon G., M{\"{u}}ller K.-R., Samek W.}:
\newblock {"What is relevant in a text document?": An interpretable machine
  learning approach}.
\newblock \emph{PLOS ONE 12}, 8 (Aug 2017).

\bibitem[AL20]{Ahn2020}
\textsc{{Ahn} Y., {Lin} Y.}:
\newblock Fairsight: Visual analytics for fairness in decision making.
\newblock \emph{IEEE Transactions on Visualization and Computer Graphics 26}, 1
  (Jan 2020), 1086--1095.

\bibitem[ALMK16]{Angwin2016}
\textsc{Angwin J., Larson J., Mattu S., Kirchner L.}:
\newblock Machine bias.
\newblock https://www.propublica.org/article/, 2016.

\bibitem[AMA{\etalchar{*}}14]{Alsallakh2014}
\textsc{Alsallakh B., Micallef L., Aigner W., Hauser H., Miksch S., Rodgers
  P.}:
\newblock {Visualizing Sets and Set-typed Data: State-of-the-Art and Future
  Challenges}.
\newblock In \emph{Eurovis STAR Reports} (2014), The Eurographics Association.

\bibitem[AR15]{Aubry2015}
\textsc{Aubry M., Russell B.~C.}:
\newblock {Understanding Deep Features with Computer-Generated Imagery}.
\newblock In \emph{2015 IEEE International Conference on Computer Vision
  (ICCV)} (Dec 2015), IEEE, pp.~2875--2883.

\bibitem[Ber10]{Bertin2010}
\textsc{Bertin J.}:
\newblock \emph{Semiology of {{Graphics}}}, 2nd~ed.
\newblock {ESRI Press}, {Redlands, CA}, 2010.

\bibitem[Bre01]{Breiman2001}
\textsc{Breiman L.}:
\newblock {Random Forests}.
\newblock \emph{Machine Learning 45}, 1 (2001), 5--32.

\bibitem[CBS{\etalchar{*}}16]{Choi2016-retain}
\textsc{Choi E., Bahadori M.~T., Sun J., Kulas J., Schuetz A., Stewart W.}:
\newblock Retain: An interpretable predictive model for healthcare using
  reverse time attention mechanism.
\newblock In \emph{Advances in Neural Information Processing Systems} (2016),
  pp.~3504--3512.

\bibitem[CD19]{Cavallo2019}
\textsc{Cavallo M., Demiralp {\c{C}}.}:
\newblock Clustrophile 2: Guided visual clustering analysis.
\newblock \emph{IEEE transactions on visualization and computer graphics 25}, 1
  (2019), 267--276.

\bibitem[CEH{\etalchar{*}}19]{Cabrera2019}
\textsc{Cabrera A.~A., Epperson W., Hohman F., Kahng M., Morgenstern J., Chau
  D.~H.}:
\newblock Fairvis: Visual analytics for discovering intersectional bias in
  machine learning.
\newblock \emph{IEEE Conference on Visual Analytics Science and Technology
  (VAST)} (2019).

\bibitem[CGM{\etalchar{*}}17]{Ceneda2017}
\textsc{Ceneda D., Gschwandtner T., May T., Miksch S., Schulz H.-J., Streit M.,
  Tominski C.}:
\newblock {Characterizing Guidance in Visual Analytics}.
\newblock \emph{IEEE Transactions on Visualization and Computer Graphics 23}, 1
  (Jan 2017), 111--120.

\bibitem[CHH{\etalchar{*}}19]{Cashman2019}
\textsc{Cashman D., Humayoun S.~R., Heimerl F., Park K., Das S., Thompson J.,
  Saket B., Mosca A., Stasko J., Endert A., Gleicher M., Chang R.}:
\newblock A user-based visual analytics workflow for exploratory model
  analysis.
\newblock \emph{Computer Graphics Forum 38}, 3 (2019), 185--199.

\bibitem[CLG{\etalchar{*}}15]{Caruana2015}
\textsc{Caruana R., Lou Y., Gehrke J., Koch P., Sturm M., Elhadad N.}:
\newblock {Intelligible Models for HealthCare}.
\newblock In \emph{Proceedings of the 21th ACM SIGKDD International Conference
  on Knowledge Discovery and Data Mining - KDD '15} (New York, New York, USA,
  2015), ACM Press, pp.~1721--1730.

\bibitem[Cra96]{Craven1996thesis}
\textsc{Craven M.}:
\newblock \emph{{Extracting Comprehensible Models from Trained Neural
  Networks}}.
\newblock {Ph. D. dissertation}, University of Wisconsin - Madison, 1996.

\bibitem[CS68]{Chen1968}
\textsc{Chen C.~F., Shieh L.~S.}:
\newblock {A novel approach to linear model simplification}.
\newblock \emph{International Journal of Control 8}, 6 (Dec 1968), 561--570.

\bibitem[CS96]{Craven96-nips}
\textsc{Craven M.~W., Shavlik J.~W.}:
\newblock {Extracting tree-structured representations of trained neural
  networks}.
\newblock In \emph{Advances in Neural Information Processing Systems} (1996),
  vol.~8, pp.~24--30.

\bibitem[DWOB20]{Dasgupta2020}
\textsc{{Dasgupta} A., {Wang} H., {O'Brien} N., {Burrows} S.}:
\newblock Separating the wheat from the chaff: Comparative visual cues for
  transparent diagnostics of competing models.
\newblock \emph{IEEE Transactions on Visualization and Computer Graphics 26}, 1
  (Jan 2020), 1043--1053.

\bibitem[EF10]{Elmqvist2010}
\textsc{Elmqvist N., Fekete J.-D.}:
\newblock Hierarchical aggregation for information visualization: Overview,
  techniques, and design guidelines.
\newblock \emph{IEEE transactions on visualization and computer graphics 16}, 3
  (2010), 439--54.

\bibitem[FSV{\etalchar{*}}19]{Friedler2019}
\textsc{Friedler S.~A., Scheidegger C., Venkatasubramanian S., Choudhary S.,
  Hamilton E.~P., Roth D.}:
\newblock A comparative study of fairness-enhancing interventions in machine
  learning.
\newblock In \emph{Proceedings of the Conference on Fairness, Accountability,
  and Transparency} (New York, NY, USA, 2019), FAT* '19, ACM, pp.~329--338.

\bibitem[GHG{\etalchar{*}}19]{Gil19}
\textsc{Gil Y., Honaker J., Gupta S., Ma Y., D'Orazio V., Garijo D., Gadewar
  S., Yang Q., Jahanshad N.}:
\newblock Towards human-guided machine learning.
\newblock In \emph{Proceedings of the 24th {{International Conference}} on
  {{Intelligent User Interfaces}} - {{IUI}} '19} ({Marina del Ray, California},
  2019), {ACM Press}, pp.~614--624.

\bibitem[{Gle}13]{Gleicher2013}
\textsc{{Gleicher} M.}:
\newblock Explainers: Expert explorations with crafted projections.
\newblock \emph{IEEE Transactions on Visualization and Computer Graphics 19},
  12 (Dec 2013), 2042--2051.

\bibitem[Gle16]{Gleicher2016}
\textsc{Gleicher M.}:
\newblock {A Framework for Considering Comprehensibility in Modeling}.
\newblock \emph{Big Data 4}, 2 (Jun 2016), 75--88.

\bibitem[Gle18]{Gleicher2018}
\textsc{Gleicher M.}:
\newblock {Considerations for Visualizing Comparison}.
\newblock \emph{IEEE Transactions on Visualization and Computer Graphics 24}, 1
  (Jan 2018), 413--423.

\bibitem[HG18]{Heimerl2018}
\textsc{Heimerl F., Gleicher M.}:
\newblock Interactive analysis of word vector embeddings.
\newblock \emph{Computer Graphics Forum 37}, 3 (2018), 253--265.

\bibitem[HPB{\etalchar{*}}14]{Henelius2014}
\textsc{Henelius A., Puolam{\"{a}}ki K., Bostr{\"{o}}m H., Asker L., Papapetrou
  P.}:
\newblock {A peek into the black box: exploring classifiers by randomization}.
\newblock \emph{Data Mining and Knowledge Discovery 28}, 5-6 (Sep 2014),
  1503--1529.

\bibitem[HPRP20]{Hohman2020}
\textsc{{Hohman} F., {Park} H., {Robinson} C., {Polo Chau} D.~H.}:
\newblock Summit: Scaling deep learning interpretability by visualizing
  activation and attribution summarizations.
\newblock \emph{IEEE Transactions on Visualization and Computer Graphics 26}, 1
  (Jan 2020), 1096--1106.

\bibitem[IK11]{Ishizaki2011}
\textsc{Ishizaki S., Kaufer D.}:
\newblock {DocuScope: Computer-aided rhetorical analysis}.
\newblock In \emph{Applied Natural Language Processing and Content Analysis:
  Advances in Identification, Investigation, and Resolution}, McCarthy P.,
  Boonthum C., (Eds.). IGI Global, 2011.

\bibitem[KDS{\etalchar{*}}17]{Krause2017}
\textsc{Krause J., Dasgupta A., Swartz J., Aphinyanaphongs Y., Bertini E.}:
\newblock A workflow for visual diagnostics of binary classifiers using
  instance-level explanations.
\newblock In \emph{2017 IEEE Conference on Visual Analytics Science and
  Technology (VAST)} (2017), IEEE, pp.~162--172.

\bibitem[KEV{\etalchar{*}}17]{Kwon2017}
\textsc{Kwon B.~C., Eysenbach B., Verma J., Ng K., DeFilippi C., Stewart W.~F.,
  Perer A.}:
\newblock {Clustervision: Visual Supervision of Unsupervised Clustering}.
\newblock \emph{IEEE Transactions on Visualization and Computer Graphics}
  (2017), 1--1.

\bibitem[KLTH10]{Kapoor2010}
\textsc{Kapoor A., Lee B., Tan D., Horvitz E.}:
\newblock {Interactive optimization for steering machine classification}.
\newblock In \emph{Proceedings of the 28th international conference on Human
  factors in computing systems - CHI '10} (New York, New York, USA, Apr 2010),
  ACM Press, p.~1343.

\bibitem[LCGH13]{Lou2013}
\textsc{Lou Y., Caruana R., Gehrke J., Hooker G.}:
\newblock {Accurate intelligible models with pairwise interactions}.
\newblock In \emph{Proceedings of the 19th ACM SIGKDD international conference
  on Knowledge discovery and data mining - KDD '13} (New York, New York, USA,
  2013), ACM Press, p.~623.

\bibitem[Lip16]{Lipton2016}
\textsc{Lipton Z.~C.}:
\newblock {The Mythos of Model Interpretability}.
\newblock \emph{arXiv preprint arXiv:1606.03490} (Jun 2016), 1606.03490.

\bibitem[LSC{\etalchar{*}}18]{MLiu2017a}
\textsc{{Liu} M., {Shi} J., {Cao} K., {Zhu} J., {Liu} S.}:
\newblock Analyzing the training processes of deep generative models.
\newblock \emph{IEEE Transactions on Visualization and Computer Graphics 24}, 1
  (Jan 2018), 77--87.

\bibitem[LSC19]{Lee2019}
\textsc{Lee K., Sood A., Craven M.}:
\newblock {Understanding Learned Models by Identifying Important Features at
  the Right Resolution}.
\newblock In \emph{AAAI} (Nov 2019).

\bibitem[LSL{\etalchar{*}}17]{Liu2017}
\textsc{{Liu} M., {Shi} J., {Li} Z., {Li} C., {Zhu} J., {Liu} S.}:
\newblock Towards better analysis of deep convolutional neural networks.
\newblock \emph{IEEE Transactions on Visualization and Computer Graphics 23}, 1
  (Jan 2017), 91--100.

\bibitem[MBVV07]{Martens2007}
\textsc{Martens D., Baesens B., {Van Gestel} T., Vanthienen J.}:
\newblock {Comprehensible credit scoring models using rule extraction from
  support vector machines}.
\newblock \emph{European Journal of Operational Research 183}, 3 (Dec 2007),
  1466--1476.

\bibitem[MCZ{\etalchar{*}}17]{Ming2017}
\textsc{{Ming} Y., {Cao} S., {Zhang} R., {Li} Z., {Chen} Y., {Song} Y., {Qu}
  H.}:
\newblock Understanding hidden memories of recurrent neural networks.
\newblock In \emph{2017 IEEE Conference on Visual Analytics Science and
  Technology (VAST)} (Oct 2017), pp.~13--24.

\bibitem[MLMP18]{Muhlbacher2018}
\textsc{Muhlbacher T., Linhardt L., Moller T., Piringer H.}:
\newblock {TreePOD: Sensitivity-Aware Selection of Pareto-Optimal Decision
  Trees}.
\newblock \emph{IEEE Transactions on Visualization and Computer Graphics 24}, 1
  (Jan 2018), 174--183.

\bibitem[MMD{\etalchar{*}}19]{Murugesan2019}
\textsc{{Murugesan} S., {Malik} S., {Du} F., {Koh} E., {Lai} T.~M.}:
\newblock Deepcompare: Visual and interactive comparison of deep learning model
  performance.
\newblock \emph{IEEE Computer Graphics and Applications 39}, 5 (Sep 2019),
  47--59.

\bibitem[MP13]{muhlbacher2013partition}
\textsc{M\"uhlbacher T., Piringer H.}:
\newblock {A Partition-Based Framework for Building and Validating Regression
  Models}.
\newblock \emph{IEEE Transactions on Visualization and Computer Graphics 19},
  12 (Dec 2013), 1962--1971.

\bibitem[MQB19]{Ming2019}
\textsc{Ming Y., Qu H., Bertini E.}:
\newblock Rulematrix: Visualizing and understanding classifiers with rules.
\newblock \emph{IEEE transactions on visualization and computer graphics 25}, 1
  (2019), 342--352.

\bibitem[MV16]{Mahendran2016}
\textsc{Mahendran A., Vedaldi A.}:
\newblock {Visualizing Deep Convolutional Neural Networks Using Natural
  Pre-images}.
\newblock \emph{International Journal of Computer Vision 120}, 3 (Dec 2016),
  233--255.

\bibitem[MXC{\etalchar{*}}20]{Ming2020}
\textsc{{Ming} Y., {Xu} P., {Cheng} F., {Qu} H., {Ren} L.}:
\newblock Protosteer: Steering deep sequence model with prototypes.
\newblock \emph{IEEE Transactions on Visualization and Computer Graphics 26}, 1
  (Jan 2020), 238--248.

\bibitem[MXLM20]{Ma2020}
\textsc{{Ma} Y., {Xie} T., {Li} J., {Maciejewski} R.}:
\newblock Explaining vulnerabilities to adversarial machine learning through
  visual analytics.
\newblock \emph{IEEE Transactions on Visualization and Computer Graphics 26}, 1
  (Jan 2020), 1075--1085.

\bibitem[OJ02]{Olden2002}
\textsc{Olden J.~D., Jackson D.~A.}:
\newblock {Illuminating the “black box”: a randomization approach for
  understanding variable contributions in artificial neural networks}.
\newblock \emph{Ecological Modelling 154}, 1-2 (Aug 2002), 135--150.

\bibitem[OW16]{Obermann2016}
\textsc{Obermann L., Waack S.}:
\newblock {Interpretable Multiclass Models for Corporate Credit Rating Capable
  of Expressing Doubt}.
\newblock \emph{Frontiers in Applied Mathematics and Statistics 2} (Oct 2016).

\bibitem[Pow11]{Powers2011}
\textsc{Powers D.}:
\newblock {Evaluation: From Precision, Recall and F-Measure to ROC,
  Informedness, Markedness and Correllation}.
\newblock \emph{Journal of Machine Learning Technologies 2}, 1 (2011), 37--63.

\bibitem[RAL{\etalchar{*}}17]{Ren2017}
\textsc{{Ren} D., {Amershi} S., {Lee} B., {Suh} J., {Williams} J.~D.}:
\newblock Squares: Supporting interactive performance analysis for multiclass
  classifiers.
\newblock \emph{IEEE Transactions on Visualization and Computer Graphics 23}, 1
  (Jan 2017), 61--70.

\bibitem[RFFT17]{Rauber2017}
\textsc{Rauber P.~E., Fadel S.~G., Falcao A.~X., Telea A.~C.}:
\newblock {Visualizing the Hidden Activity of Artificial Neural Networks}.
\newblock \emph{IEEE Transactions on Visualization and Computer Graphics 23}, 1
  (Jan 2017), 101--110.

\bibitem[RSG16]{Ribeiro2016}
\textsc{Ribeiro M.~T., Singh S., Guestrin C.}:
\newblock {``Why Should I Trust You?'': Explaining the Predictions of Any
  Classifier}.
\newblock In \emph{Proceedings of the 22nd ACM SIGKDD International Conference
  on Knowledge Discovery and Data Mining - KDD '16} (New York, New York, USA,
  Feb 2016), ACM Press, pp.~1135--1144.

\bibitem[RW18]{Rodrigues2018}
\textsc{{Rodrigues} N., {Weiskopf} D.}:
\newblock Nonlinear dot plots.
\newblock \emph{IEEE Transactions on Visualization and Computer Graphics 24}, 1
  (Jan 2018), 616--625.

\bibitem[SAMG14]{Sarikaya2014}
\textsc{Sarikaya A., Albers D., Mitchell J., Gleicher M.}:
\newblock {Visualizing Validation of Protein Surface Classifiers}.
\newblock \emph{Computer Graphics Forum 33}, 3 (Jun 2014), 171--180.

\bibitem[SCF{\etalchar{*}}19]{Visus}
\textsc{Santos A., Castelo S., Felix C., Ono J.~P., Yu B., Hong S., Silva
  C.~T., Bertini E., Freire J.}:
\newblock Visus: {{An Interactive System}} for {{Automatic Machine Learning
  Model Building}} and {{Curation}}.
\newblock In \emph{2019 {{Workshop}} on {{Human}}-{{In}}-the-{{Loop Data
  Analytics}} ({{HILDA}}'19)} (Jul 2019).

\bibitem[SDV{\etalchar{*}}16]{Selvaraju2016a}
\textsc{Selvaraju R.~R., Das A., Vedantam R., Cogswell M., Parikh D., Batra
  D.}:
\newblock {Grad-CAM: Why did you say that?}
\newblock \emph{arXiv preprint arXiv:1611.07450} (Nov 2016).

\bibitem[SDW09]{Slingsby2009}
\textsc{Slingsby A., Dykes J., Wood J.}:
\newblock Configuring {{Hierarchical Layouts}} to {{Address Research
  Questions}}.
\newblock \emph{IEEE Transactions on Visualization and Computer Graphics 15}, 6
  (Nov 2009), 977--984.

\bibitem[SGB{\etalchar{*}}19]{Strobelt2019}
\textsc{Strobelt H., Gehrmann S., Behrisch M., Perer A., Pfister H., Rush
  A.~M.}:
\newblock Seq2seq-vis: A visual debugging tool for sequence-to-sequence models.
\newblock \emph{IEEE transactions on visualization and computer graphics 25}, 1
  (2019), 353--363.

\bibitem[SGK17]{Shrikumar2017}
\textsc{Shrikumar A., Greenside P., Kundaje A.}:
\newblock {Learning Important Features Through Propagating Activation
  Differences}.
\newblock \emph{PMLR 70} (Apr 2017), 3145--3153.

\bibitem[SGPR18]{Strobelt1018}
\textsc{{Strobelt} H., {Gehrmann} S., {Pfister} H., {Rush} A.~M.}:
\newblock Lstmvis: A tool for visual analysis of hidden state dynamics in
  recurrent neural networks.
\newblock \emph{IEEE Transactions on Visualization and Computer Graphics 24}, 1
  (Jan 2018), 667--676.

\bibitem[{\v{S}}K14]{Strumbelj2014}
\textsc{{\v{S}}trumbelj E., Kononenko I.}:
\newblock {Explaining prediction models and individual predictions with feature
  contributions}.
\newblock \emph{Knowledge and Information Systems 41}, 3 (Dec 2014), 647--665.

\bibitem[SSSE20]{Spinner2020}
\textsc{{Spinner} T., {Schlegel} U., {Schäfer} H., {El-Assady} M.}:
\newblock explainer: A visual analytics framework for interactive and
  explainable machine learning.
\newblock \emph{IEEE Transactions on Visualization and Computer Graphics 26}, 1
  (Jan 2020), 1064--1074.

\bibitem[SSSG16]{Szafir2016textDNA}
\textsc{Szafir D.~A., Stuffer D., Sohail Y., Gleicher M.}:
\newblock {TextDNA: Visualizing Word Usage with Configurable Colorfields}.
\newblock \emph{Computer Graphics Forum 35}, 3 (Jun 2016), 421--430.

\bibitem[STY17]{Sundararajan2017}
\textsc{Sundararajan M., Taly A., Yan Q.}:
\newblock Axiomatic attribution for deep networks.
\newblock In \emph{Proceedings of the 34th International Conference on Machine
  Learning-Volume 70} (2017), JMLR.org, pp.~3319--3328.

\bibitem[SVC{\etalchar{*}}10]{Steyerberg2010}
\textsc{Steyerberg E.~W., Vickers A.~J., Cook N.~R., Gerds T., Gonen M.,
  Obuchowski N., Pencina M.~J., Kattan M.~W.}:
\newblock {Assessing the performance of prediction models: a framework for
  traditional and novel measures.}
\newblock \emph{Epidemiology (Cambridge, Mass.) 21}, 1 (Jan 2010), 128--38.

\bibitem[SVZ13]{Simonyan2013}
\textsc{Simonyan K., Vedaldi A., Zisserman A.}:
\newblock {Deep Inside Convolutional Networks: Visualising Image Classification
  Models and Saliency Maps}.
\newblock \emph{arXiv preprint arXiv:1312.6034} (Dec 2013).

\bibitem[Sza18]{szafirModelingColorDifference2018}
\textsc{Szafir D.~A.}:
\newblock Modeling {{Color Difference}} for {{Visualization Design}}.
\newblock \emph{IEEE Transactions on Visualization and Computer Graphics 24}, 1
  (Jan 2018), 392--401.

\bibitem[TKDB17]{Tamagnini2017}
\textsc{Tamagnini P., Krause J., Dasgupta A., Bertini E.}:
\newblock {Interpreting Black-Box Classifiers Using Instance-Level Visual
  Explanations}.
\newblock In \emph{Proceedings of the 2nd Workshop on Human-In-the-Loop Data
  Analytics - HILDA'17} (New York, New York, USA, 2017), ACM Press, pp.~1--6.

\bibitem[TLKT09]{Talbot2009}
\textsc{Talbot J., Lee B., Kapoor A., Tan D.~S.}:
\newblock {EnsembleMatrix}.
\newblock In \emph{Proceedings of the 27th international conference on Human
  factors in computing systems - CHI 09} (New York, New York, USA, Apr 2009),
  ACM Press, p.~1283.

\bibitem[{Wal}84]{Wallis1984}
\textsc{{Wallis, Jerold W and Shortliffe} E.~H.}:
\newblock {Customized explanations using causal knowledge}.
\newblock In \emph{Rule-Based Expert Systems: The MYCIN Experiments of the
  Stanford Heuristic Programming Project}. Addison-Wesley, Reading, MA, 1984,
  pp.~371--388.

\bibitem[Wei80]{Weiner1980}
\textsc{Weiner J.}:
\newblock {BLAH, a system which explains its reasoning}.
\newblock \emph{Artificial Intelligence 15}, 1-2 (Nov 1980), 19--48.

\bibitem[WFH11]{Witten2011}
\textsc{Witten I.~H., Frank E., Hall M.~A.}:
\newblock \emph{{Data Mining: Practical Machine Learning Tools and Techniques,
  3e}}.
\newblock Morgan Kaufmann, Burlington, 2011.

\bibitem[WH10]{Witmore2010}
\textsc{Witmore M., Hope J.}:
\newblock {The Hundredth Psalm to the Tune of “Green Sleeves”: Digital
  Approaches to Shakespeare's Language of Genre}.
\newblock \emph{Shakespeare Quarterly 61}, 3 (2010), 357--390.

\bibitem[Wil05]{GrammarOfGraphics}
\textsc{Wilkinson L.}:
\newblock \emph{The {{Grammar}} of {{Graphics}}, Second Edition}, 2nd~ed.
\newblock Statistics and {{Computing}}. {Springer-Verlag}, {New York}, 2005.

\bibitem[WMJ{\etalchar{*}}19]{ATMSeer}
\textsc{Wang Q., Ming Y., Jin Z., Shen Q., Liu D., Smith M.~J., Veeramachaneni
  K., Qu H.}:
\newblock {{ATMSeer}}: {{Increasing Transparency}} and {{Controllability}} in
  {{Automated Machine Learning}}.
\newblock In \emph{Proceedings of the 2019 {{CHI Conference}} on {{Human
  Factors}} in {{Computing Systems}} - {{CHI}} '19} ({Glasgow, Scotland Uk},
  2019), {ACM Press}, pp.~1--12.

\bibitem[WPB{\etalchar{*}}20]{Wexler2020}
\textsc{{Wexler} J., {Pushkarna} M., {Bolukbasi} T., {Wattenberg} M., {Viégas}
  F., {Wilson} J.}:
\newblock The what-if tool: Interactive probing of machine learning models.
\newblock \emph{IEEE Transactions on Visualization and Computer Graphics 26}, 1
  (Jan 2020), 56--65.

\bibitem[WR15]{Wang2015-fallingRuleLists}
\textsc{Wang F., Rudin C.}:
\newblock {Falling Rule Lists}.
\newblock In \emph{18th International Conference on Artificial Intelligence and
  Statistics (AISTATS)} (Nov 2015).

\bibitem[WSW{\etalchar{*}}17]{Wongsuphasawat2017}
\textsc{Wongsuphasawat K., Smilkov D., Wexler J., Wilson J., Mane D., Fritz D.,
  Krishnan D., Viegas F.~B., Wattenberg M.}:
\newblock Visualizing dataflow graphs of deep learning models in tensorflow.
\newblock \emph{IEEE Transactions on Visualization and Computer Graphics 24}, 1
  (2017), 1--12.

\bibitem[YCN{\etalchar{*}}15]{Yosinski2015}
\textsc{Yosinski J., Clune J., Nguyen A., Fuchs T., Lipson H.}:
\newblock {Understanding Neural Networks Through Deep Visualization}.
\newblock In \emph{Deep Learning Workshop, 31 st International Conference on
  Machine Learning} (2015).

\bibitem[YXX{\etalchar{*}}19]{Ye2019}
\textsc{Ye X., Xiang S., Xia J., Wu J., Chen Y., Lu S.}:
\newblock Interactive correction of mislabeled training data.
\newblock \emph{IEEE Conference on Visual Analytics Science and Technology
  (VAST)} (2019).

\bibitem[ZF14]{Zeiler2014}
\textsc{Zeiler M., Fergus R.}:
\newblock {Visualizing and Understanding Convolutional Networks}.
\newblock In \emph{ECCV 2014} (Cham, 2014), Fleet D., Pajdla T., Schiele B.,
  Tuytelaars T., (Eds.), vol.~8689 of \emph{Lecture Notes in Computer Science},
  Springer International Publishing, pp.~818--833.

\bibitem[ZWM{\etalchar{*}}19]{Zhang2019}
\textsc{Zhang J., Wang Y., Molino P., Li L., Ebert D.~S.}:
\newblock Manifold: A model-agnostic framework for interpretation and diagnosis
  of machine learning models.
\newblock \emph{IEEE transactions on visualization and computer graphics 25}, 1
  (2019), 364--373.

\end{thebibliography}
\bibliographystyle{eg-alpha}



\end{document}